\documentclass[conference]{IEEEtran}

\usepackage{graphicx}
\usepackage{balance} 
\usepackage{amsmath,amssymb,amsfonts}
\usepackage{textcomp}
\usepackage{xcolor}
\usepackage{booktabs} 
\usepackage{algpseudocode}
\usepackage{color}
\usepackage{xcolor}

\usepackage{soul}
\usepackage{url}
\usepackage[linesnumbered,ruled]{algorithm2e}
\usepackage{hhline}
\usepackage{multirow}
\usepackage{makecell}
\usepackage{caption}
\usepackage[caption=false,font=footnotesize,labelfont=sf,textfont=sf]{subfig}

\usepackage{tikz}

\usepackage[flushleft]{threeparttable}
\usepackage{framed}
\usepackage{multicol}
\usepackage{rotating} 

\renewenvironment{framed}[1][\hsize]
   {\MakeFramed{\hsize#1\advance\hsize-\width \FrameRestore}}%
   {\endMakeFramed}

\usepackage{lipsum}

\usepackage{array}
\newtheorem{definition}{Definition}
\newtheorem{theorem}{Theorem}

\begin{document}

\title{Accelerating Forward and Backward Private Searchable Encryption Using Trusted Execution}

% IEEE author section
\author{\IEEEauthorblockN{Viet Vo${}^\dag$${}^\ddag$, Shangqi Lai${}^\dag$${}^\ddag$, Xingliang Yuan${}^\dag$, Shi-Feng Sun${}^\dag$${}^\ddag$, Surya Nepal${}^\ddag$, and Joseph K. Liu${}^\dag$}

\IEEEauthorblockA{${}^\dag$Monash University, Australia; 
${}^\ddag$Data61, CSIRO, Australia;\\
\{viet.vo, shangqi.lai, xingliang.yuan, shifeng.sun, joseph.liu\}@monash.edu; surya.nepal@data61.csiro.au}
}

%\IEEEoverridecommandlockouts
%\makeatletter\def\@IEEEpubidpullup{6.5\baselineskip}\makeatother
%\IEEEpubid{\parbox{\columnwidth}{
%    Network and Distributed Systems Security (NDSS) Symposium 2020\\
%    23-26 February 2020, San Diego, CA, USA\\
%    ISBN 1-891562-61-4\\
%    https://dx.doi.org/10.14722/ndss.2020.23xxx\\
%    www.ndss-symposium.org
%}
%\hspace{\columnsep}\makebox[\columnwidth]{}}
% make the title area
\maketitle

\begin{abstract}

Searchable encryption (SE) is one of the key enablers for building encrypted databases. It allows a cloud server to search over encrypted data without decryption. Dynamic SE additionally includes data addition and deletion operations to enrich the functions of encrypted databases. Recent attacks exploiting the leakage in dynamic operations drive the rapid development of SE schemes revealing less information while performing updates; they are also known as forward and backward private SE. Newly added data is no longer linkable to queries issued before, and deleted data is no longer searchable in queries issued later. However, those advanced SE schemes reduce the efficiency of SE, especially in the communication cost between the client and server. In this paper, we resort to the hardware-assisted solution, aka Intel SGX, to ease the above bottleneck. Our key idea is to leverage SGX to take over most tasks of the client, i.e., tracking keyword states along with data addition and caching deleted data. However, handling large datasets is non-trivial due to the I/O and memory constraints of SGX. We further develop batch data processing and state compression techniques to reduce the communication overhead between the SGX and untrusted server and minimise the memory footprint within the enclave. We conduct a comprehensive set of evaluations on both synthetic and real-world datasets, which confirm that our designs outperform the prior art.

\end{abstract}

\section{Introduction}

Searchable encryption (SE)~\cite{SoWa00,CurtmolaGKO06} is designed to enable a user to outsource her data to remote servers securely while preserving search functionalities. It is considered as the most promising solution to build encrypted databases defending against data breaches. Generic solutions like fully homomorphic encryption, multi-party computation, and oblivious RAM (ORAM) achieve strong security but introducing considerable computational and communication overhead. Property-preserving encryption like deterministic encryption and order-preserving/revealing encryption is efficient and legacy compatible in databases, but those solutions are not secure in practice~\cite{Bindschaedler18}. The reasonable security and performance tradeoff brought by SE continuously drives the rapid development of new SE schemes with more functionalities~\cite{Zuo18} and improved security~\cite{Bost16,RaphaelBO17,LaiPSL18}. 

In~\cite{CashGP15}, Cash et al. introduced the concept of active attacks against dynamic SE; the leakage in data update operations can be exploited to compromise the claimed security of SE. After that, Zhang et al.~\cite{ZhangKP16} proposed the first instantiation of active attacks called file-injection attacks through the exploitation of the leakage in data addition. This work raises a natural question: whether a dynamic SE scheme with less leakage can be designed to mitigate existing and even prevent prospective active attacks. To address this question, forward and backward private SE schemes~\cite{Bost16,RaphaelBO17,SunYLS18,GharehChamani18} have drawn much attention recently. 

In dynamic SE, the notion of forward privacy means that the linkability between newly added data and previously issued search queries should be hidden against the server, and the notion of backward privacy means that the linkability between deleted data and search queries after deletion should be hidden. To achieve higher security for SE, the efficiency of SE is compromised. Existing forward and backward private SE schemes~\cite{RaphaelBO17,SunYLS18,GharehChamani18} introduce large overhead in storage and computation at both client and server, and/or increase the client-server interaction. In order to maintain the efficiency of SE, an alternative approach is to employ the  hardware-assisted solution, i.e., Intel SGX, where native code and data can be executed in a trusted and isolated execution environment. 
%
%SGX provides I/O communication interface to support trusted execution from a given application. They are \textit{ecall}-invoking a function within SGX from the application, and \textit{ocall}-invoking a function defined in the application from SGX.
%
Recent work in ORAM powered by SGX~\cite{Mishra18} demonstrates that SGX can be treated as a delegate of clients, so as to ease the overhead of client storage and computation, and reduce the communication cost between the client and server. 

Amjad et al.~\cite{Amjad19} proposed the first forward and backward private SE schemes using SGX. 
As generic ORAM or ORAM-like data structures can natively be adapted to achieve the strongest forward and backward privacy in SE (i.e., Type-I~\cite{RaphaelBO17}), one of their schemes is built from ORAM, where data addition and deletion are completely oblivious to the server~\cite{Amjad19}. It is noteworthy that such an approach could still be inefficient due to the high I/O complexity between SGX and server. Like prior forward and backward private SE studies, Amjad et al. also proposed an efficient scheme (i.e., Type-II~\cite{RaphaelBO17}) that trades security for higher efficiency named \textsf{Bunker-B}~\cite{Amjad19}. Timestamps of update operations will not be exposed, while the rounds of interaction between the SGX and server can be reduced. 
In this work, we are interested in designs with forward and Type-II backward privacy due to its practical balance between security and efficiency.

Unfortunately, only the theoretical construction of \textsf{Bunker-B} is given in~\cite{Amjad19}, and we observe that it is not scalable, especially when handling large datasets. First, deletion operations are realised via insertion operations, which will (a) incur large communication costs between the SGX and server, i.e., the number of \textit{ocall}s scales with the number of deletions, and (b) increase search latency, because all deleted data needs to be retrieved, decrypted, and filtered out from the search results. 
Second, re-encryption is adopted after each search for forward and backward privacy, which will also incur long search latency and affect the performance of other concurrent queries. The reason is that if deleted documents are only a small portion of the matched results, most of the results (non-deleted ones) need to be re-encrypted and re-inserted to the database.
More detailed analysis can be found in Section~\ref{sec:evaluation}.

% Unfortunately, only the theoretical construction of \textsf{Bunker} is given in~\cite{Amjad19}, and we observe that it is still far from being practical, especially when handling large datasets.
% %
% First, deletion operations are realised via insertion operations over keyword-document pairs, which will incur large communication costs, i.e., the number of \textit{ocall}s scales with the number of keyword-document pairs. 
% %
% We find that \textsf{Bunker} needs 10$\times$ more I/O operations for addition and 30$\times$ more for deletion.
% %
% As presented in the evaluation, its addition and deletion performance is 2$\times$ and 3$\times$ slower than our schemes respectively.
% %
% In addition, it would requires Bunker-B more $m$ ocall for physically deleting $m$ documents.
% %Second, client still needs to maintain keyword states for forward privacy, which will  increase the communication cost between the client and SGX in search and addition, because search and addition tokens are generated by the client. Third, 
% % We found that Bunker-is x2 slower than ours during insertion throughput, and deletion/throughputs.
% % 
% Second, re-encryption is adopted after each search for backward privacy, which will incur long search latency. The reason is that if deleted documents are only a small portion of the matched results, most of the results (non-deleted ones) need to be re-encrypted and inserted to the database. 
% %
% We observe that \textsf{Bunker-B} can be 2x slower than our schemes if the deletion portion is set to 25\%.
% %

To avoid the potential performance bottleneck introduced by SGX, in this paper, we devise forward and backward private SE schemes from a simple yet effective approach. Our idea is to leverage the SGX enclave to fully act as the client. The enclave will cache both the keyword state and the deletions, so as to reduce the communication cost and roundtrips between the SGX and server in search, addition, and deletion operations, and make the client almost free in computation and storage. Furthermore, we propose several optimisations to accelerate the performance, including batch document processing, state compression via Bloom filter, and memory efficient implementation. 

\vspace{2pt}
\noindent \textbf{Contributions}: Our contributions in this paper can be summarised as follows:

\begin{itemize}

\item We design and implement two forward and backward private SE schemes, named \textsf{SGX-SE1} and \textsf{SGX-SE2}. By using SGX, the  communication cost between the client and server of achieving forward and backward privacy in SE is significantly reduced. 

\item Both \textsf{SGX-SE1} and \textsf{SGX-SE2} leverage the SGX enclave to carefully track keyword states and document deletions, in order to minimise the communication overhead between the SGX and untrusted memory. In particular, \textsf{SGX-SE2} is an optimised version of \textsf{SGX-SE1} by employing Bloom filter to compress the information of deletions, which speeds up the search operations and  boosts the capacity of batch processing in addition and deletion. 

\item We formalise the security model of our schemes and perform security analysis accordingly. 

\item We conduct comprehensive evaluations on both synthetic and real-world datasets. Our experiments show that the latest art \textsf{Bunker-B} takes $10\times$ more \textit{ecall/ocalls} than our schemes \textsf{SGX-SE1} and \textsf{SGX-SE2} when inserting $10^6$ documents. Even more, \textsf{Bunker-B} needs $30\times$ \textit{ecall/ocalls} when deleting $25\%$ of the above documents.  W.r.t. search latency,  \textsf{SGX-SE1} and \textsf{SGX-SE2} are $30\%$ and $2\times$ faster than \textsf{Bunker-B}, respectively.     

%after inserting $10^6$ documents and caching $2.5\times 10^5$ deleted documents
\end{itemize}

%\vspace{2pt}
%\noindent \textbf{Organisation}: We discuss some related works in Section~\ref{sec:relatedworks}. Section~\ref{sec:background} presents the background of SGX and dynamic SE. Section~\ref{sec:proposed_schemes} details our system architecture, threat models, design intuition, and presents our proposed schemes \textsf{SGX-SE1} and \textsf{SGX-SE2} with their security analysis. In  Section~\ref{sec:evaluation}, we evaluate the schemes and compare them with the prior art. Section~\ref{sec:con} concludes the paper.

\vspace{-5pt}
\section{Related Work}
\label{sec:relatedworks}

%In this section, we overview searchable encryption (SE) and different approaches that improve search efficiency in SE. We then present    significant works that rely on trusted execution environments (TEEs) to do so. Finally, we refer to advanced security notions of SE when supporting document updates.
%\vspace{2pt}
\noindent \textbf{Searchable encryption}: Song et al.~\cite{SoWa00} presented the first searchable encryption (SE) to enable search over encrypted documents. After that, Curtmola et al.~\cite{CurtmolaGKO06} and Kamara et al.~\cite{Kamara12} formalised the security definitions for static and dynamic SE, respectively, and proposed  schemes with sublinear search time. 
%\vspace{2pt}
%\noindent \textbf{Search efficiency}: 
Since SSE was formalised, a long line of studies has been proposed to improve query efficiency~\cite{Cash14} and support expressive queries~\cite{Zuo18}. 

\vspace{2pt}
\noindent \textbf{Forward and backward privacy in SE}: In dynamic SE, forward privacy means data additions do not reveal their associations to any query made in the past, and deleted documents cannot be accessed via any post queries. Forward privacy has been studied widely to mitigate file-injection attacks~\cite{ZhangKP16,StefanovPS14,Bost16}. Backward privacy has received less attention~\cite{RaphaelBO17,SunYLS18,GharehChamani18,Zuo19}. There are three types of backward privacy from Type-I to Type-III in the descending order of security. However, strong backward private (Type-I and Type-II) schemes are known to be inefficient in computation and communication overhead, as shown in~\cite{RaphaelBO17,GharehChamani18}. 

\vspace{2pt}
\noindent \textbf{Encrypted search with trusted execution}: Another line of research in this field~\cite{Amjad19,Priebe18,Mishra18,Fuhry17} is to leverage trusted execution environment (TEE). In general, TEE such as Intel SGX can reduce the network roundtrips between the client and server and enrich the database functions in the encrypted domain. Fuhry et al.~\cite{Fuhry17} proposed \textsf{HardIDX} that organises database index in a $B^+$-tree structure and utilises enclave to traverse a subset of the tree nodes to do searches. 
%The scheme achieves search time complexity at the logarithm in the size of index, but it does not support dynamic update operations. 
%
Later, Mishra et al.~\cite{Mishra18} designed a doubly-oblivious SE scheme that supports inserts and deletes, named \textsf{Oblix}. In this scheme, one oblivious data index resides in the enclave to map the search index of each keyword to a location in another oblivious structure located in untrusted memory. %However, the performance of their implementation on large databases is less efficient due to the fact of using ORAM. 
Regarding SE, Borges et al.~\cite{Borges18} migrated secure computation to the enclave to improve the search efficiency of SE boolean queries. When two or more keywords are queried, the result set can be unionised or intersected within the enclave. Note that this work focuses on a different problem with ours.
Very recently, Amjad et al.~\cite{Amjad19} proposed three schemes to enable single-keyword query with different search leakage (i.e., information that the server can learn about the query and data). However, the practical performance of these schemes has not been investigated.  
Meanwhile, Ren et al.~\cite{Ren20} proposed a volume-hiding range query scheme via SGX.

%Papers from DB community:
%1. Wu, Songrui, et al. "Servedb: Secure, verifiable, and efficient range queries on outsourced database." 2019 IEEE 35th International Conference on Data Engineering (ICDE). IEEE, 2019.

%2. Demertzis, Ioannis, Rajdeep Talapatra, and Charalampos Papamanthou. "Efficient searchable encryption through compression." Proceedings of the VLDB Endowment 11.11 (2018): 1729-1741.

%3. Demertzis, Ioannis, and Charalampos Papamanthou. "Fast searchable encryption with tunable locality." Proceedings of the 2017 ACM International Conference on Management of Data. ACM, 2017.

%4. SecEQP: A Secure and Efficient Scheme for SkNN Query Problem over Encrypted Geodata on Cloud .Xinyu Lei, Alex X. Liu, Rui Li, Guan-Hua Tu .In Proceedings of the 35th IEEE International Conference on Data Engineering (ICDE), Macau SAR, April 2019.

%5. Fast Range Query Processing with Strong Privacy Protection for Cloud Computing. Rui Li, Alex X. Liu, Ann L. Wang, and Bezawada Bruhadeshwar. In Proceedings of the 40th International Conference on Very Large Data Bases (VLDB), Vol. 7, No. 14, Hangzhou, China, September, 2014.

%6. ShieldStore: Shielded In-memory Key-value Storage with SGX, eurosys 2019

\vspace{-5pt}
\section{Background}
\label{sec:background}
\vspace{-5pt}
\subsection{Intel SGX}

Intel SGX is a set of x86 instructions designed for improving the security of application code and data.
On SGX-enabled platforms, ones need to partition the application into both trusted part and untrusted part.
The trusted part, dubbed enclave, is located in a dedicated memory portion of physical RAM with strong protection enforced by SGX.
The untrusted part is executed as an ordinary process and can invoke the enclave only through the well-defined interface, named \textit{ecall}, while the enclave can encrypt clear data and send to untrusted code via the interface named \textit{ocall}. 
Furthermore, decryption and integrity checks are performed when the data is loaded inside the enclave.
All other software, including OS, privileged software, hypervisor, and firmware cannot access the enclave's memory.
%
%In that way, outside applications (e.g. malicious software) cannot learn the plaintext. 
%
%SGX's enclave is constructed with very limited $128$ MB cache memory size. 
%
%That memory is used for both SGX metadata and the enclave itself. 
%
The actual memory for storing data in the enclave is only up to $96$ MB. 
Above that, SGX will automatically apply page swapping. % with proper integrity and confidentiality guarantees when allocating more than $96$ MB. 
SGX also has a remote attestation feature that allows to verify the creation of enclaves on a remote server and to create a secure communication channel to the enclaves. 
%
%During the creation, initial code and data are loaded into the enclave for measurement. 
%
%Then, the enclave attests itself to the remote provider via authentication checking. 
%
%After that, an encrypted communication channel can be established between two parties. 
%
%Secrets such as credentials or sensitive data can be provisioned directly to the enclave.

%While SGX brings a promising efficiency solution to SE, it leaks the data access pattern since the untrusted OS still manages enclave's resources and observes the enclaves behaviour. Then, the OS can trace the enclave's code and data accesses at the granularity of pages. 
%
There has been a lot of existing SGX side-channel attacks such as hardware side-channels~\cite{Yarom14}, cache timing~\cite{Brasser17}, and page-fault attacks~\cite{Shinde16}. 
However, we are also aware that the security in future SGX versions will be improved~\cite{Brasser19}. % by both hardware and software based countermeasures. 
%
%For instance, Dr. SGX~\cite{Brasser19} provided new data randomisation strategies to defence against side-channel attacks that target data access patterns. 
%
%Cloak~\cite{Gruss17} implements new hardware transaction memory techniques to detect and prevent adversarial observation of cache misses.
%
%Sanctum~\cite{Costan16} and Varys~\cite{Oleksenko18} provide new verifiable hardware/software extensions to mitigate the attacks.%page-fault attacks.  

\subsection{Dynamic Searchable Symmetric Encryption}
\vspace{-2pt}
\label{subsec:dsse}
In this section, we briefly overview dynamic SE and the notion of forward and backward privacy in dynamic SE.
%
%More details of forward and backward privacy are in~\cite{Bost16,RaphaelBO17}.
%
Following the verbatim in~\cite{Bost16,RaphaelBO17}, let \textsf{DB} represent a database of  documents, and each document \textsf{doc} with a unique identifier \textit{id} is a variable-length set of unique keywords. We use \textsf{DB}($w$) to present the set of documents where keyword $w$ occurs. The total number of keyword-document pairs is denoted by $N$, $W$ is the total number of distinct keywords in \textsf{DB}. All $N$ keyword-document pairs are stored in an index $M_I$, which is a dictionary structure mapping each unique keyword $w$ to a list of matching documents in  \textsf{DB}($w$). The encrypted database, named \textsf{EDB} is a collection of encrypted documents. A dynamic SE scheme $\Sigma=(\textsf{Setup}, \textsf{Search}, \textsf{Update})$ consists of three protocols between a client and a server as follows:

\noindent$\textsf{Setup}(1^\lambda,\textsf{DB})$: The protocol inputs a security parameter $\lambda$ and outputs a secret key $K$,  a state $ST$ for the client, and an encrypted database \textsf{EDB}. %that will be sent to the server. 

\noindent$\textsf{Search}(K,w,ST;\textsf{EDB})$: The protocol allows to query $w$ based on the state $ST$, the secret key $K$ and the state $ST$ from the client, and the encrypted database \textsf{EDB} from the server. After that, it outputs the search result $Res$.%. matching $w$.

\noindent$\textsf{Update}(K,(\textsf{op},\textsf{in}),ST;\textsf{EDB})$: The protocol takes $K$, $ST$, an input \textsf{in} associated with an operation \textsf{op} from the client, and \textsf{EDB}, where $\textsf{op} \in \{add,del\}$ and \textsf{in} consists of a document identifier \textit{id} and a set of keywords in that document. Then, the protocol inserts or removes \textsf{in} from \textsf{EDB} upon \textsf{op}.

Giving a list of queries $Q$ sent by the client, the server records the timestamps $u$ for every query with $Q=\{q: q=(u,w)\ \textnormal{or}~ (u,\textsf{op},\textsf{in})\}$. 
Following the verbatim from~\cite{Bost16,RaphaelBO17}, we let $\textsf{TimeDB}(w)$ be the access pattern which consists of the non-deleted documents \textit{currently} matching $w$ and the timestamps of inserting them to the database. Formally,
\begin{equation} 
\label{eq1}
\notag
\begin{split}
\textsf{TimeDB}(w) = \{(u,id): &(u,add,(w,id)) \in Q \\
            & \textnormal{and}~\forall u^\prime, (u^\prime,del,(w,id)) \notin Q \}
\end{split}
\end{equation}
\noindent and let $\textsf{Updates}(w)$ be the list of timestamps of updates:
\begin{equation} 
\label{eq2}
\notag
\begin{split}
\textsf{Updates}(w)=\{u: (u,\textsf{op},(w,id)) \in Q\}
\end{split}
\end{equation}

\noindent There are two security properties based on the leakage function of dynamic SE~\cite{RaphaelBO17}.
The \textit{forward privacy} ensures that each update leaks no information about the keyword that was queried in the past and currently is in the document to be updated. 
%
%More specifically, when a client wants to update \textsf{EDB} with a keyword $w$ and a document identifier $id$, the server must not learn whether $w$ was searched in the past or not.
%
The \textit{backward privacy} guarantees that when a keyword-document pair $(w,id)$ is added and then deleted, subsequent searches on $w$ do not reveal $id$. 
%
%Ideally, a backward-private SSE scheme should leak nothing about the deletions, and at least not reveal the identifiers of deleted documents~\cite{RaphaelBO17}. 
%
%However, as formalised in~\cite{RaphaelBO17}, $id$ should be revealed as a result if a search on $w$ is performed after inserting $(w,id)$ but before removing it. 
%
%According to~\cite{RaphaelBO17}, the \textit{backward privacy} does not apply to searching the same keyword-document pairs happens between the addition and the deletion. 
%
There are three types of \textit{backward privacy} with varying levels of leakages from Type-I to Type-III introduced in~\cite{RaphaelBO17}. 
Type-I backward privacy is the most secure. 
It only reveals what time the current (non-deleted) documents matching to $w$ added (i.e.,\textsf{TimeDB}(w)). 
Type-II additionally leaks what time updates on $w$ made, presented as  $\{\textsf{TimeDB}(w),\textsf{Updates}(w)\}$. 
In a less secure manner, Type-III inherits the leakage of Type-II and additionally reveals which addition updates cancel which deletion updates.
%
%However, our study focuses on  Type-II that leaks and reveals when the updates on $w$ happen.
%
%Current Type-I schemes \textsf{Moneta}~\cite{RaphaelBO17} and \textsf{Orion}~\cite{GharehChamani18}, which relies on ORAM-based~\cite{Stefanov12}, are inefficient due to ORAM's overhead. 
%

Current Type-II schemes \textsf{Fides}~\cite{RaphaelBO17} and \textsf{Mitra}~\cite{GharehChamani18} require multiple roundtrips and high communication cost, while \textsf{Horus}~\cite{GharehChamani18} relies on Path-ORAM.%~\cite{Stefanov13}.
%
%In a less secure manner, Type-III schemes \textsf{Janus}~\cite{RaphaelBO17} and \textsf{Janus++}~\cite{SunYLS18} only require one roundtrip as a tradeoff between communication cost and security guarantees. 
 %
 Until recently, Amjad et al.~\cite{Amjad19} proposed three  SGX-supported schemes, including the Type-I scheme \textsf{Fort}, Type-II scheme \textsf{Bunker-B}, and Type-III scheme \textsf{Bunker-A}. 
 However, \textsf{Fort} requires an oblivious map (OMAP) similar to the one in \textsf{Orion}~\cite{GharehChamani18} to do the update, causing high computation overhead. 
 \textsf{Bunker-A}~\cite{Amjad19} improves the update computation, but it downgrades the security guarantees. 
 In contrast, \textsf{Bunker-B} is designed with a good tradeoff in computation/communication cost and security guarantees. 
 We will later compare the performance of \textsf{Bunker-B} with our schemes in Section~\ref{sec:evaluation}.

\section{Our Proposed Schemes}
\label{sec:proposed_schemes}

We present the overview of our proposed schemes, as shown in Fig.1. 
%
%The system is designed in a general way such that it can apply any of previous forward and backward SGX-supported SSE schemes, such as \textsf{Bunker-A}, \textsf{Bunker-B}, \textsf{Fort}~\cite{Amjad19}, and our \textsf{SGX-SE1} and \textsf{SGX-SE2} as well. 
%
After that, we detail our scheme design intuition by analysing previous SGX-supported schemes~\cite{Amjad19} in terms of communication/computation overhead and then highlight our technical solution. 
Finally, we present \textsf{SGX-SE1} and \textsf{SGX-SE2} with corresponding protocols.% to resolve the recognised limitations.

\subsection{System Overview}
The design involves three entities: the client (who is the data owner and therefore trusted), the untrusted server, and the trusted SGX enclave within the server. The system flow involves 9 steps.
\begin{figure}
\centering
\includegraphics[width=0.45\textwidth,height=4.2cm]{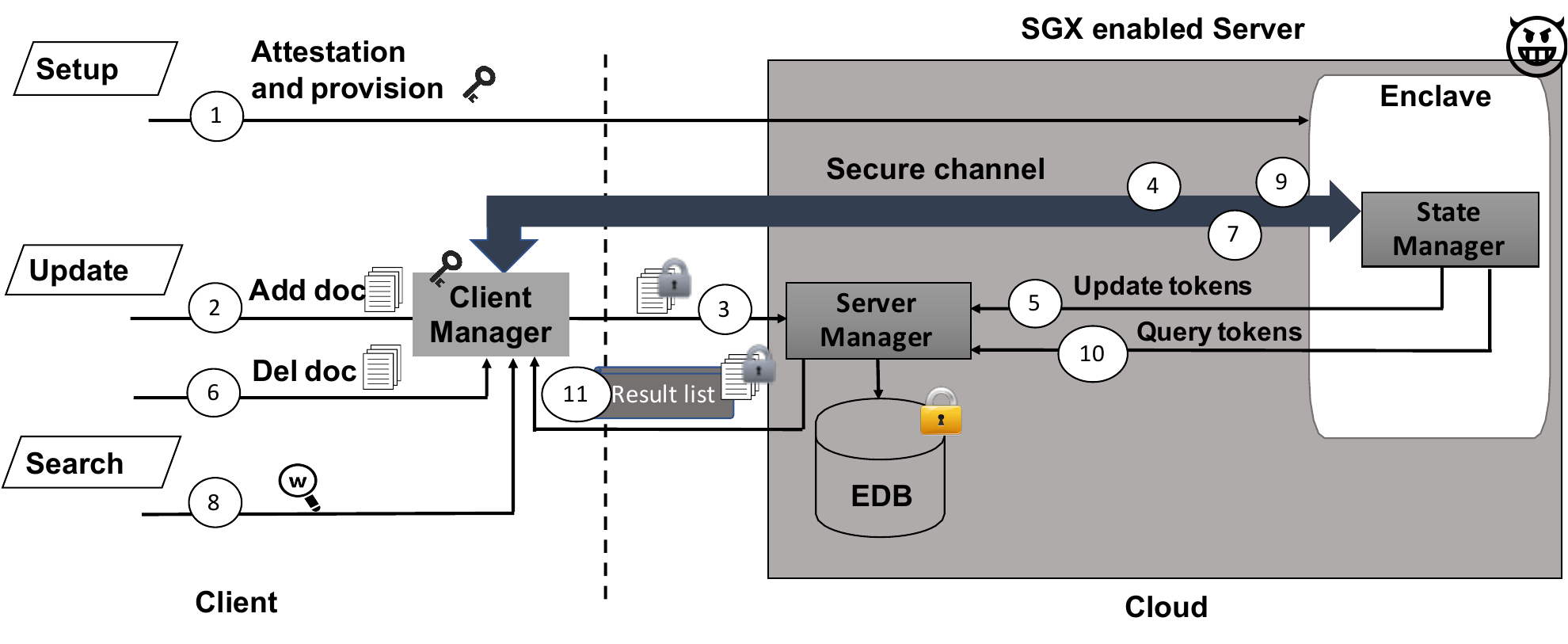} 
%\vspace{-20pt}
\caption{High level design}
\vspace{-15pt}
\label{fig:system}
\end{figure}

At step 1, the client uses the SGX attestation feature to authenticate the enclave and establish a secure channel with the  enclave. 
The client then provisions a secret key $K$ to the enclave through this channel. 
This completes the \textsf{Setup} protocol of our proposed protocol. 
Note that this operation does not deploy any \textsf{EDB} to the server as in dynamic SE schemes~\cite{RaphaelBO17}. 
Instead, we consider that the client outsources documents to the server via \textsf{Update} operations later.

At step 2, giving a document with a unique identifier $id$, the \textit{Client Manager} encrypts the document with the key $K$ and sends the encrypted version of the document to the \textit{Server Manager} (see step 3). 
The encrypted version with its $id$ is then inserted to \textsf{EDB}. 
After that, the \textit{Client Manager} sends the original document to the \textit{State Manager} located in the enclave via the secure channel (see step 4). 
At this step, the \textit{State Manager} performs cryptographic operations to generate update tokens that will be sent to \textit{Server Manager} (see step 5). 
The tokens are used to update the encrypted index of dynamic SE located in the \textit{Server Manager}. 
Note that traditional dynamic SE schemes~\cite{Cash14,RaphaelBO17,SunYLS18} often consider \textsf{EDB} as the underlying encrypted index of dynamic SE, and omit the data structure storing encrypted documents. 
Here, we locate them separately to avoid that ambiguity, i.e., the index of dynamic SE $M_I$ is located in \textit{Server Manager}, and encrypted documents reside in \textsf{EDB} as an encrypted document repository, respectively.
To delete a document with a given $id$ (step 6), the \textit{Client Manager} directly sends the document $id$ to the \textit{State Manager} (see step 7). 
%If the system is deploying forward and backward secure SSE schemes such as \textsf{Bunker-B}, there are additional steps between \textit{State Manager} and \textit{Server Manager} to retrieve and update $M_I$.

At step 8, the client wants to search documents matching a given query keyword $w$. 
The \textit{Client Manager} will send the keyword $w$ to the \textit{State Manager} (see step 9). 
Then, the \textit{State Manager} computes query tokens and excludes the tokens for deleted documents according to the deletion information from step 6.
Later, the \textit{State Manager} sends them to the \textit{Server Manager} (in step 10). 
The \textit{Server Manager} will search over the received tokens and return the list of encrypted matching documents back to the \textit{Client Manager}. 
At that stage, the encrypted documents are decrypted with $K$.

\subsection{Assumptions and Threat Models}

%As we develop the DSSE system based on SGX's security guarantee, we focus on \textit{forward and backward privacy} of DSSE and exclude all SGX's vulnerabilities like many other SGX applications~\cite{Huayi19,Mishra18,Orenbach17}.
%
%\vspace{2pt}
\noindent \textbf{Our Assumptions with Intel SGX}: We assume that SGX behaves correctly, (i.e., there are no hardware bugs or backdoors), and the preset code and data inside the enclave are protected. 
Also, the communication between the client and the enclave relies on the secure channel created during SGX attestation. 
%From SSE's point of view, the client relies on the enclave to process plaintext and execute cryptographic operations. 
%
%Then, enclave will call outside untrusted programs (i.e. \textit{Server Manager}) via \textit{ocall}s to store encrypted data. 
%
Like many other SGX applications~\cite{Huayi19,Mishra18}, side-channel attacks~\cite{Yarom14,Brasser17,Shinde16} against SGX are out of our scope.
Denial-of-service (DoS) attacks are also out of our focus, i.e., the enclave is always available whenever the client invokes or queries. 
Finally, we assume that all the used cryptographic primitives and libraries of SGX are trusted.

\vspace{5pt}
\noindent \textbf{Threat Models}: Like existing work~\cite{Fuhry17,Amjad19}, we consider a semi-honest but powerful attacker at the server-side.
Although the attacker will not deviate from the protocol, he/she can gain full access over software stack outside of the enclave, OS and hypervisor, as well as hardware components in the server except for the processor package. 
 In particular, the attacker can observe memory addresses and (encrypted) data on the memory bus, in memory, or in \textsf{EDB} to generate data access patterns. 
 Additionally, the attacker can log the time when these memory manipulations happen.
 The goal of the attacker is to learn extra information about the encrypted database from the leakage both revealed by hardware and the leakage function defined in section~\ref{sec:security}.

%\subsection{\textsf{Bunker-B}}
%\label{sec:bunkerB}

%The \textsf{Update} and \textsf{Search} protocols of \textsf{Bunker-B} are summarily in Algorithm~\ref{alg:BunkerB}.
%
%The scheme requires $O(1)$ update computation complexity and $a_w$ update \textit{ocall}s.   
%
%For each keyword/doc pair, \textsf{Bunker-B} requires enclave to generate tokens for addition and deletion and uses the generated tokens to update $M_I$ on the server (line 5 in Algorithm~\ref{alg:BunkerB}).
%
%In \textsf{Search} operation, the scheme requires $O(a_w)$ and involves  (i.e., $a_w$) roundtrips. Filtering non-deleted ids and re-encryption technique is applied to them for every search (see lines 21-26 in Algorithm~\ref{alg:BunkerB}).

\begin{algorithm}[!t]
\DontPrintSemicolon

\textsf{Update}$(op,in):$ // $op\in \{add,del\},~in=(w,id)$\\ \vspace{3pt}
    
\Indp 
    Client retrieves $st_w=(version,count)$ from $st$;\\
    Send $(w,version,count,op,id)$ to enclave;\\
    Client updates $st_w=(version,count+1)$ to $st$;\\

    Enclave generates an update token $u_{tk}=(u,v)$:
    $u:=F_{K_1}(w||version||count+1)$
    $v:=Enc(K_2, id||op)$\\
    Enclave sends $u_{tk}$ to the server;\\
    Server receives $u_{tk}=(u,v)$ from the enclave;\\
    Server updates the map $M_I[u]=v$ \vspace{3pt}\\
\Indm    

\textsf{Search}$(w):$ \vspace{3pt}   

\Indp
    Client retrieves $st_w=(version,count)$ from $st$;\\
    Client outputs $(w,version,count)$ to enclave $st$;\\
    Client updates $st_w=(version+1,count)$ to $st$;\\
    Enclave receives $(w,version,count)$ from client;\\
    Enclave generates query tokens $q_{tk}=(u_1,\ldots,u_i,\ldots,u_{count})$, where :\\
\Indp \Indp
    $u_i:=F_{K_1}(w||version||i)$\\
\Indm\Indm   
    Enclave sends $q_{tk}$ to the server;\\
    Server returns to the enclave with the list $L=\{(u_1,v_1),\ldots,(u_c,v_c)\}$;\\
    Server deletes all pairs in the $L$ from $M_I$;\\
    Enclave filters non-deleted $id$s with $R=\{id: \nexists (id,op=del)\in L\}$;\\
    Enclave returns $R$ to the client;\\
    Enclave resets $count=1$ and re-encrypts $R$ with
    
    \textbf{foreach} $id \in R$:
        Generate a new token \\
\Indp \Indp
        $u:=F_{K_1}(w||version+1||count)$\\
        $v:=Enc(K_2, id||op=add)$\\
        Send $(u,v)$ to the server to update $M_I$;\\
        Enclave increase $count+=1;$\\
\Indm\Indm    
 
\SetAlCapNameFnt{\small }
\SetAlCapFnt{\small }
\small 
\caption{\textsf{Bunker-B}~\cite{Amjad19}: \textsf{Update} and \textsf{Search} protocols}
\label{alg:BunkerB}
\end{algorithm}

%\begin{sidewaystable*}
\begin{table*}[t]
\small
\centering
  \caption{Comparison with previous SGX-supported schemes. $N$, $D$, and $W$ denote the total number of keyword/document pairs, total number of documents, and total number of keywords, respectively. $d$ presents the number of deleted documents. $n_w$ is the number of (current, non-deleted) documents containing $w$, $a_w$ is the total number of entries (including addition and deletion updates) performed on $w$, $d_w$ denotes the number of deletions performed on $w$. $r$ is the predefined number of necessary dummy entries to be inserted in oblivious operations. $v_d$ denotes the vector of a Bloom filter to check the membership of $\#d$ documents.}
  \label{tab:comparison}
  %\vspace{-10pt}
  \begin{threeparttable}[!t]
  \tabcolsep 0.02in
  \begin{tabular}{|c|c|c|c|c|c|c|c|c|c|}
    \hline
    \multirow{3}*{SGX Schemes} & \multicolumn{4}{c|}{Communication between enclave and server} & \multicolumn{2}{c|}{Enclave Computation} & \multirow{2}{*}{Client} & \multirow{2}{*}{Enclave} & \multirow{2}{*}{BP} \\
\cline{2-7}
     &\#Search rounds& \multirow{2}{*}{Search} & \#Update & \multirow{2}{*}{Update} & \multirow{2}{*}{Search} & \multirow{2}{*}{Update} &  & & \\
     &(\textit{ecall} + \textit{ocall}) & & \textit{ocalls} & & & & Storage & Storage & Type\\
    \hline
    \textsf{Fort} \cite{Amjad19}   & $a_w$ & $O(n_w)$ & $(a_w + r)$ & $O(1)$ & $O(n_w + \sum_w d_w)$ & $O(\log^2 N)$ & $O(W\log D)$ & -- & I\\
    \hline
    \hline
    \textsf{Bunker-B} \cite{Amjad19}  & $a_w$ & $O(n_w)$ & $a_w$ & $O(1)$ & $O(a_w)$ & $O(1)$ & $O(W\log D)$ & -- & II \\
    \hline
    \textsf{SGX-SE1} & ($d$ + $d_w$)$\star$ & $O(n_w)$ & $n_w$\dag & $O(1)$ & $O(n_w + d)\ddag$ & $O(1)$ & -- & $O(W\log D + d)$ & II \\
    \hline    
    \textsf{SGX-SE2} & ($d_w$)$\star$  & $O(n_w)$ & $n_w$\dag & $O(1)$ & $O(n_w + v_d)$ & $O(1)$ & -- & $O(W\log D)\star\star$ & II \\
    %\textsf{Moneta} \cite{DBLP:conf/ccs/BostMO17}   & 3 & $\tilde{O}(a_w \log N + \log^3 N)$ & $\tilde{O}(\log^3 N)$ & $\tilde{O}(a_w \log N + \log^3 N)$ & $\tilde{O}(\log^2 N)$ & $O(1)$ & Type-I \\
    %\hline
    %\textsf{Orion} \cite{DBLP:conf/ccs/ChamaniPPJ18}  & $O(\log N)$ & $O(n_w \log^2 N)$ & $O(\log^2 N)$ & $O(n_w \log^2 N)$ & $O(\log^2 N)$ & $O(1)$ & Type-I \\
    \hline
    \hline
    \textsf{Bunker-A} \cite{Amjad19}  & $a_w$ & $O(n_w)$ &  $a_w$ & $O(1)$ & $O(a_w)$& $O(1)$ & $O(W\log D)$ & -- & III \\    
    %\textsf{Fides} \cite{DBLP:conf/ccs/BostMO17}  & 2 & $O(a_w)$ &  $O(1)$ & $O(a_w)$& $O(1)$ & $O(W\log D)$ & Type-II \\
    %\hline
    %\textsf{Mitra} \cite{DBLP:conf/ccs/ChamaniPPJ18}  & 2 & $O(a_w)$ &  $O(1)$ & $O(a_w)$& $O(1)$ & $O(W\log D)$ &Type-II \\
    %\hline
    %$\textsf{Aura}_{\textsf{I}}$/$\textsf{Aura}_{\textsf{II}}$  & 1 & $O(n_w)$ &  $O(1)$ & $O(n_w)$& $O(1)$ &$O(Wd)\dag$  &  Type-II \\
    %\hline
    %\hline

    %\hline
    %\textsf{Diana$_{\textsf{del}}$} \cite{DBLP:conf/ccs/BostMO17}  & 2 & $O(n_w+d_w\log a_w)$ &  $O(1)$ & $O(a_w)$& $O(\log a_w)$ & $O(W\log D)$ &Type-III \\
    %\hline
    %\texttt{Janus} \cite{DBLP:conf/ccs/BostMO17}  & 1 & $O(n_w)$ &  $O(1)$ & $O(n_wd_w)$& $O(1)$ & $O(W\log D)$ & Type-III  \\
    %\hline
    %\textsf{Horus} \cite{DBLP:conf/ccs/ChamaniPPJ18}  & $O(\log d_w)$ & $O(n_w\log d_w \log N)$ &  $O(\log^2 N)$ & $O(n_w\log d_w \log N)$ & $O(\log^2 N)$ & $O(W\log D)$ &Type-III \\
    %\hline
    %\texttt{Janus++} \cite{DBLP:conf/ccs/SunYLSSVN18} & 1 & $O(n_w)$ & $O(1)$ & $O(n_w d)$& $O(d)$ & $O(W\log D)$ &Type-III \\
    %\hline
    %$\textsf{LattiSE}$ & 1 &  $O(n_w)$ &  $O(1)$ & $O(n_wd_w)$& $O(1)$ & $O(W\log D)$ & Type-III \\
    \hline
  \end{tabular}
 \begin{tablenotes}
\item $\star$: The complexity also requires $n_w$ \textit{ocall}s (one-way trip) when sending query tokens to the server. 
\item $\star\star$: The complexity also requires the size of a configurable Bloom filter vector. 
\item $\dag$: We note that the number of update \textit{ocall}s is $n_w$ if the update is addition. Otherwise, deletion updates do not take any \textit{ocall}s.
\item $\ddag$: If there is no deletion updates between two searches on different $w$, $d$ is cancelled. Then, the complexity is only $O(n_w)$.
\end{tablenotes}
\end{threeparttable}
\end{table*}
%\end{sidewaystable*}

\subsection {Design Intuition}
\label{subsec:design}

As mentioned, Amjad et al.~\cite{Amjad19} proposed three backward private SGX-supported schemes: the Type-I scheme \textsf{Fort}, Type-II scheme \textsf{Bunker-B}, and Type-III scheme \textsf{Bunker-A}. 
The performance and security overview of these schemes can be found in Table~\ref{tab:comparison}.
The table demonstrates the computation and communication cost for \textsf{update} and \textsf{search} among SGX-supported backward-private schemes.
In \textsf{update}, \textsf{Fort} requires  $(a_w + r)$ \textit{ocall}s and  $O(log^2N)$  computation complexity.
The \textsf{search} operation of \textsf{Fort} requires  $a_w$ roundtrips between the enclave and the server since the enclave needs to retrieve all the labels associated with $w$ before discarding deleted labels retrieved previously in \textsf{deletion} updates. 
\textsf{Fort} is the most secure while still relying on ORAM and thus we exclude it in this work due to its overhead.
As shown, \textsf{Bunker-B} has $O(1)$ update computation complexity and $a_w$ update \textit{ocall}s. However, it causes high computation complexity $O(a_w)$ and involves a large number of roundtrips (i.e., $a_w$) during the search. 
\textsf{Bunker-A} does not perform re-encryption and re-insertion after search and thus only achieves Type-III backward privacy. However, it still treats deletion as insertion, just like \textsf{Bunker-B}. %Therefore, \textsf{Bunker-A} is also inefficient in data deletion. 
Therefore, we only analyse the limitations of \textsf{Bunker-B} as follows.
%Clearly, the schemes do not require the interaction between the client and the server during \textsf{search} operations due to the trusted execution of the enclave, as analysed in that previous work. However, the communication rounds between the enclave and the server, and the enclave's computation cost in these schemes are still expensive. The efficiency of these schemes can be found in Table~\ref{tab:comparison}.

\vspace{5pt}
\noindent \textbf{Performance Analysis of Prior Work}:  
The \textsf{Update} and \textsf{Search} protocols of \textsf{Bunker-B} are summarily presented in Algorithm~\ref{alg:BunkerB}. 
As shown, \textsf{Bunker-B} only requires $O(1)$ update computation complexity and $a_w$ update \textit{ocall}s.   
For each $(w,id)$, \textsf{Bunker-B} lets the enclave follow the same routine to generate tokens for addition and deletion and uses the generated tokens to update $M_I$ on the server ( line 5 in Algorithm~\ref{alg:BunkerB}).
However, it causes high computation complexity $O(a_w)$ and involves a large number of roundtrips (i.e., $a_w$) during the search. 
In the \textsf{Search} protocol, the core idea of \textsf{Bunker-B} is to let the enclave read all records (associated with $add$ or $del$) in $M_I$ corresponding to the keyword.
Then, the enclave decrypts them and filters deleted \textit{id}s based on the operation.
After query, the enclave re-encrypts non-deleted ids and sends the newly generated tokens to the server for updates. 
These steps are summarised in lines 21-26 in Algorithm~\ref{alg:BunkerB}. 
We have implemented \textsf{Bunker-B} (see Section~\ref{sec:evaluation}) and found that the scheme also has other limitations in practice as follows:

\textit{Intensive Ecall/Ocall Usage}: Giving a document \textsf{doc} with an identifier $id$ and $M$ unique keywords to the server, \textsf{Bunker-B} repeatedly performs the \textsf{Update} protocol by using $M$ \textit{ecall}s and then the same number of \textit{ocall}s to insert tokens to the index map $M_I$. 
It indicates that the number of \textit{ecall/ocall} for \textsf{Bunker-B} is linear to the keyword-document pairs for updates.
In practice, a dataset can include a large number of keyword-document pairs ($>10^7$).
As a result, \textsf{Bunker-B} takes $12\mu$s to insert one $(w,id)$ pair, and  $2.36\times  10^7$ \textit{ecall/ocalls} to insert $10^6$ documents to the database. 
Similarly, deleting a \textsf{doc} in \textsf{Bunker-B} is the same as the addition, with the exception that the tokens contain $op=del$. 
Experimentally, \textsf{Bunker-B} takes $1.98\times 10^8$ $ecall/ocall$s to delete $2.5\times 10^5$ documents. 
The practical performance of Bunker-B can be found in Section~\ref{sec:evaluation}. 
We also note that \textsf{Bunker-B} only supports deletion updates on the index map $M_I$ without considering deleting real documents~\cite{Amjad19}. 
%
%To do so, \textsf{Bunker-B} will need additional $d_w$ \textit{ocalls} to request the server to delete $d$ documents containing $w$. 
%This deletion overhead is not mentioned in the previous work.

\textit{Search Latency}: The re-encryption on non-deleted \textit{id}s per search makes \textsf{Bunker-B} inefficient. In particular, when the number of those $id$s is large and the deleted ones is a small portion (adding $10^6$ documents and deleting $25\%$ documents), \textsf{Bunker-B} takes $3.2$s to query a keyword (see Section~\ref{sec:evaluation}).

%\textit{Memory Bottleneck}: During \textsf{search}, the enclave loads all non-deleted and deleted \textit{id}s back to its memory to filter non-deleted ones. That union operation is also expensive, it may exceed the limited memory of the enclave when the number of updates on $w$ is very large. In particular, \textsf{Bunker-B} exceeds the memory limit (128 MB) when the size of query result larger than $2.5\times 10^5$.

\vspace{5pt}
\noindent \textbf{Technical Highlights}: Motivated by the limitations of \textsf{Bunker-B}, we design \textsf{SGX-SE1} and \textsf{SGX-SE2} that are Type-II backward private schemes with: (1) reduced number of \textit{ecall/ocall} when the client wants to add/delete a document, (2) reduced search roundtrips, and (3) accelerated enclave's  computation in search.

We achieve (1) by allowing the client to transfer the document to the enclave for document addition, instead of transferring $(w,id)$ pairs. 
This design reduces the number of \textit{ecall}s to $1$. 
We then use the enclave to store the latest states $ST$ of all keywords, where the state of a keyword $w$ is $ST[w]=count$. 
As a result, the enclave is able to generate addition tokens based on $ST$. 
Our experiments (see Section~\ref{sec:evaluation}) show that this design improves $2\times$ the addition throughput compared to \textsf{Bunker-B}. 
We note that it is negligible to store $ST$ in the enclave since it costs less than 6 MB to store the states of all keywords in the American dictionary of English\footnote{The dictionary contains about 300,000 common and obsolete keywords} (assuming each keyword state item can take up $18$ bytes in a dictionary map). 
Additionally, our scheme only requires $1$ \textit{ecall} if the client deletes a document, by transferring that document $id$ to the enclave.

W.r.t. (2), the \textsf{SGX-SE1} scheme reduces the search roundtrips between the enclave and the server to $(d + d_w)$.  
The basic idea behind \textsf{SGX-SE1} is to let the enclave cache the mapping between $w$ and the deleted document $id$s. In particular, the enclave loads and decrypts $d$ deleted documents to extract the mapping $(w,id)$. It cleans the memory after loading each deleted document to avoid the memory bottleneck. After that, the enclave needs $d_w$ roundtrips to retrieve the \textit{counter}s when the enclave filters those deleted \textit{id}s. 
\textsf{SGX-SE2} is more optimal by requiring only $d_w$ roundtrips without the need for loading $d$ deleted documents. 
To do this, \textsf{SGX-SE2} uses a Bloom filter \textit{BF} to store the mapping $(w,id)$ within the enclave. Note that the \textit{BF} can track $1.18\times10^7$ $(w,id)$ pairs with the storage cost of $34$ MB enclave memory\footnote{$1.18\times10^7$ pairs $\approx 386\times$ \texttt{Hamlet} tragedy written by William Shakespeare} with the false positive probability $P_e=10^{-4}$. 
Our experiments (see Section~\ref{sec:evaluation}) show that the search latency of \textsf{SGX-SE1} is $30\%$ faster than \textsf{Bunker-B} after inserting $10^6$ documents and caching $2.5\times 10^5$ deleted documents. Moreover, \textsf{SGX-SE2} is $2\times$ faster than \textsf{Bunker-B} for the query after deleting $25\%$ documents.

\begin{figure*}[!t]
\begin{framed}
\vspace{-10pt}
\begin{multicols}{3}

\underline{\textsf{Setup}($1^\lambda$)}\\[6pt]
\hspace*{7pt}\textit{Client:}\\
1: $k_{\Sigma}, k_f\xleftarrow{\$}\{0, 1\}^\lambda$;\\
2: Launch a remote attestation;\\
3: Establish a secure channel;\\  
4: Send $K=(k_{\Sigma}, k_f)$ to $Enclave$;\\
\hspace*{7pt}\textit{Enclave:}\\
5: Initialise maps $ST$ and $D$;\\[2pt]
6: Initialise a list $d$;\\[2pt]
7: Initialise tuples $T_1$ and $T_2$;\\[2pt]
8: Receive $K=(k_{\Sigma}, k_f)$;\\[2pt]
\hspace*{7pt}\textit{Server:}\\
9: Initialise maps $M_I$ and $M_c$;\\[2pt]
10: Initialise a repository $R$;\\[2pt]
\underline{\textsf{Update}(\textsf{op},\textsf{in})}\\[6pt]
\hspace*{7pt}\textit{Client:}\\
1: \textbf{if} $\textsf{op}=add$ \textbf{then}\\
2: \hspace*{8pt}$f \leftarrow$ \textsf{Enc}($k_f$,\textsf{doc});\\
3: \hspace*{8pt}send ($id,f$) to $Server$;\\
4: \textbf{end if}\\
5: send (\textsf{op},$id$) to $Enclave$\\ 
\hspace*{7pt}\textit{Enclave:}\\
6: \textbf{if} $\textsf{op}=add$ \textbf{then}\\
7: \hspace*{9pt}$f\leftarrow R[id]$;\\
8: \hspace*{9pt}$\{(w,id)\}\leftarrow Parse(\textsf{Dec}(k_f,f))$;\\ 
%6: \hspace*{9pt}$\{(w,id)\}\leftarrow$ $Parse$(\textsf{in});\\
9: \hspace*{9pt}\textbf{foreach} $(w,id)$ \textbf{do}\\
10:\hspace*{19pt}$k_w\parallel k_c \leftarrow F(k_\Sigma,w)$;\\
11:\hspace*{19pt}$c\leftarrow ST[w]$;\\
12:\hspace*{19pt}\textbf{if} $c= \perp$ \textbf{then} $c=-1$;\\
13:\hspace*{19pt}$c\leftarrow c+1$;\\
\columnbreak

14:\hspace*{19pt}$k_{id}\leftarrow H_1(k_w,c)$;\\
15:\hspace*{14pt}$(u,v)\leftarrow (H_2(k_w,c),{\scriptsize \textsf{Enc}(k_{id},id) }$\\
16:\hspace*{19pt}add $(u,v)$ to $T_1$;\\
17:\hspace*{14pt}$(u^\prime,v^\prime)\leftarrow (H_3(k_w,id),{\scriptsize \textsf{Enc}(k_{c},c)  }$\\
%14:\hspace*{19pt}$v^\prime\leftarrow \textsf{Enc}(k_{c},c)$;\\
18:\hspace*{19pt}add $(u^\prime,v^\prime)$ to $T_2$;\\
19:\hspace*{19pt}$ST[w]\leftarrow c$;\\
20:\hspace*{15pt}\textbf{end foreach} \\
21:\hspace*{15pt}send $(T_1,T_2)$ to $Server$;\\
22:\hspace*{15pt}reset $T_1$ and $T_2$;\\
23:\hspace*{7pt}\textbf{else} // $\textsf{op}=del$\\
24:\hspace*{17pt}add $id$ to $d$;\\
25:\hspace*{7pt}\textbf{end if}\\
\hspace*{7pt}\textit{Server:}\\
26:\hspace*{9pt}// if $\textsf{op}=add$\\
27:\hspace*{9pt}receive ($id,f$) from $Client$;\\
28:\hspace*{9pt}$R[id]\leftarrow f$; \\%// add encrypted doc\\
29:\hspace*{9pt}receive ($T_1$,$T_2$) from $Enclave$;\\
30:\hspace*{9pt}\textbf{foreach} $(u,v)$ \textbf{in} $T_1$ \textbf{do}\\
31:\hspace*{15pt} $M_I[u]\leftarrow v$;\\ 
32:\hspace*{9pt}\textbf{end foreach} \\
33:\hspace*{9pt}\textbf{foreach} $(u^\prime,v^\prime)$ \textbf{in} $T_2$ \textbf{do}\\
34:\hspace*{15pt} $M_c[u^\prime]\leftarrow v^\prime$;\\ 
35:\hspace*{9pt}\textbf{end foreach} \\
36:\hspace*{9pt}// if $\textsf{op}=del$ then do nothing\\
\underline{\textsf{Search}($w$)}\\[5pt]
\hspace*{7pt}\textit{Client:}\\
1: \hspace*{9pt}send $w$ to $Enclave$;\\
\hspace*{7pt}\textit{Enclave:}\\
2: \hspace*{9pt}$st_{w_c} \leftarrow \{\emptyset\}, Q_w \leftarrow \{\emptyset\};$\\
3: \hspace*{9pt}$k_w\parallel k_c \leftarrow F(k_\Sigma,w)$;\\
4: \hspace*{9pt}\textbf{foreach} $id_i$ \textbf{in} $d$ \textbf{do}\\
5: \hspace*{14pt}$f_i\leftarrow R[id_i]$;\\%{\scriptsize //get encrypted doc}\\ 
\columnbreak

6: \hspace*{18pt}$\textsf{doc}_i\leftarrow \textsf{Dec}(k_f,f_i)$;\\ 
7: \hspace*{18pt}\textbf{if} $w$ \textbf{in} $\textsf{doc}_i$ \textbf{then}\\
8: \hspace*{26pt}$D[w]\leftarrow  id_i \cup D[w]$;\\
9: \hspace*{26pt}delete $R[id_i]$; \\%// delete doc 
10:\hspace*{18pt}\textbf{end if} \\
%7: \hspace*{18pt}\textbf{foreach} $w_{del}$ \textbf{in} $\textsf{doc}_i$ \textbf{do}\\ 
%8: \hspace*{26pt}$D[w_{del}]\leftarrow  id_i \cup D[w_{del}]$;\\
%9: \hspace*{18pt}\textbf{end foreach} \\
%0:\hspace*{16pt}delete $R[id_i]$; // delete doc \\
11:\hspace*{9pt}\textbf{end foreach} \\
12:\hspace*{9pt}\textbf{foreach} $id$ \textbf{in} $D[w]$ \textbf{do}\\ 
13:\hspace*{15pt}$u^\prime\leftarrow H_3(k_w,id)$;\\
14:\hspace*{15pt}$v^\prime\leftarrow M_c[u^\prime]$;\\
15:\hspace*{15pt}$c\leftarrow \textsf{Dec}(k_{c},v^\prime)$;\\
16:\hspace*{15pt}$st_{w_c} \leftarrow \{c\} \cup st_{w_c}$;\\
17:\hspace*{15pt}delete $M_c[u^\prime]$;\\
18:\hspace*{9pt}\textbf{end foreach} \\
19:\hspace*{9pt}$st_{w_c} \leftarrow \{0,\ldots,ST[w]\}  \setminus st_{w_c} $\\ 
20:\hspace*{9pt}\textbf{foreach} $c$ \textbf{in} $st_{w_c}$ \textbf{do}\\
21:\hspace*{15pt}$u\leftarrow H_2(k_w,c)$;\\
22:\hspace*{15pt}$k_{id}\leftarrow H_1(k_w,c)$;\\
23:\hspace*{15pt}$Q_w \leftarrow \{(u,k_{id})\} \cup Q_w$;\\
24:\hspace*{9pt}\textbf{end foreach} \\
25:\hspace*{9pt}send $Q_w$ to $Server$;\\
26:\hspace*{9pt}delete $D[w]$;\\
\hspace*{7pt}\textit{Server:}\\
27: \hspace*{4pt}receive $Q_w$ from $Enclave$;\\
28: \hspace*{4pt}$Res \leftarrow \emptyset$; // file collection\\
29: \hspace*{4pt}\textbf{foreach} $(u_i,k_{{id}_i})$ \textbf{in} $Q_w$ \textbf{do}\\
30: \hspace*{16pt}$id_i\leftarrow \textsf{Dec}(k_{{id}_i},M_I[u_i])$;\\
31: \hspace*{16pt}$\textsf{doc}_i \leftarrow R[id_i]$;\\
32: \hspace*{16pt}add $\textsf{doc}_i$ to $Res$;\\
33: \hspace*{4pt}\textbf{end foreach}\\
34: \hspace*{4pt}send $Res$ to $Client$;\\
\hspace*{7pt}\textit{Client:}\\
35: \hspace*{4pt}decrypt $Res$ with $k_f$;\\

\end{multicols}
\vspace{-30pt}
\end{framed}
\vspace{-10pt}
\caption{Protocols in \textsf{SGX-SE1}. In \textsf{Update}, weak backward privacy (i.e., type-III) can be achieved by letting the enclave queries the deleted document from \textit{S} to update $D$. If there are no deletion updates between two searches, the enclave records the deleted $id$ to other keywords in $D[w]$.}%need to have no deletion updates between 2 searchs to cache other (keywords,id) in D[w], otherwise, run out of enclave memory.
\label{fig:X1_protocols}
\end{figure*}

W.r.t. (3), the proposed \textsf{SGX-SE1} scheme improves the search computation complexity to $O(n_w + d)$. We note that the complexity is even amortised if there is no deletion updates between a sequence of queries. The reason is that the enclave only loads $d$ document for the first query to update the mapping of all keywords in $ST$ with the deleted documents. 
Furthermore, the search computation complexity of \textsf{SGX-SE2} is only $O(n_w)$. We note that testing the membership of $d$ documents in the \textit{BF} is $v_d$ where $v$ is the vector of \textit{BF}. Our experiments (see Section~\ref{sec:evaluation}) show that \textsf{Bunker-B} takes $3.2$s for queries after inserting $10^6$ documents and deleting $25\%$ documents while \textsf{SGX-SE1} only takes $2.4$s after caching those deleted documents. In addition, \textsf{SGX-SE2} spends the least time $1.4$s, i.e., $2\times$ faster than \textsf{Bunker-B}.

%\textit{Enclave memory bottleneck}: \textsf{SGX-SE1} and {SGX-SE2} do not maintain high memory consumption during queries. In particular, they generate the query tokens for non-deleted documents and send them to the server. They do not perform any intensive the array of deleted document $ids$ between two searches. We note that The enclave in our schemes only executes ocall to ask the server searches over the M_I and then the server returns directly to the client directly. single communication direction when the client can infer non-deleted counters of  outsource the server to compute the id and send directly to the client.
%Note that id is not meaningful to the server as it was encrypted by the client. Since the clients only send $n_w$ ocalls for query storage improvement:  bost the server to faster search we do not need to maintain the duplicated like op=add,

\subsection{\textsf{SGX-SE1} Construction}

The basic idea behind \textsf{SGX-SE1} is to let the enclave store the latest states $ST$ of keywords and keeps the list $d$ of deleted document $id$s, in order to facilitate searches. Then, the enclave only loads the deleted documents for the first search between two deletion updates to update the mapping between deleted $id$s and tracked keywords. Subsequent searches between the two deletion updates do not require loading the deleted documents again. We note that the enclave clearly needs to remove $d$ after retrieving them in the first query to save the enclave's storage. Once the enclave knows the mapping between the query keyword and deleted documents, it infers the mapping of the query keyword with the rest non-deleted documents, in order to generate query tokens. After that, the server retrieves documents based on the received tokens and returns the document result list to the client. The detail protocols of \textsf{SGX-SE1} can be found in Figure~\ref{fig:X1_protocols}. We explain the protocols further as follows: 

\begin{figure*}
\begin{framed}
\vspace{-10pt}
\begin{multicols}{2}

\underline{\textsf{Setup}($1^\lambda$)}\\[6pt]
1: Performs the same \textsf{Setup} in \textsf{SGX-SE1};\\
2: \textcolor{blue}{Client inits $k_{BF}\xleftarrow{\$}\{0, 1\}^\lambda$};\\
3: \textcolor{blue}{Client sets integers $b, h$};\\  
4: \textcolor{blue}{Provisions ($k_{BF},b,h$) to Enclave};\\
5: \textcolor{blue}{Enclave selects $\{H_j^\prime\}_{j \in[h]}$ for $BF$};\\
5: \textcolor{blue}{Enclave does not maintain $D$};\\
\underline{\textsf{Update}(\textsf{op},\textsf{in})}\\[6pt]
1: Performs the same \textsf{Update} in \textsf{SGX-SE2};\\
2: \textcolor{blue}{\textbf{if} $\textsf{op}=add$ \textbf{then}}\\
3: \hspace*{13pt}\textcolor{blue}{\textbf{foreach} $(w,id)$ \textbf{do}}\\
4: \hspace*{19pt}\textcolor{blue}{\textbf{for} $j=1:h$ \textbf{do}}\\
\columnbreak

5: \hspace*{26pt}\textcolor{blue}{${\displaystyle 
h^\prime_j(w,id)\overset{\Delta}{=}H^\prime_j(k_{BF},w \parallel id)}$};\\ 
6: \hspace*{26pt}$\textcolor{blue}{{\displaystyle BF[h^\prime_j(w,id)]\leftarrow 1 }}$;\\ 
%7: \hspace*{19pt}\textcolor{blue}{\textbf{end for}} \\
%8: \hspace*{13pt}\textbf{end foreach} \\
%9: \textcolor{blue}{\textbf{end if}}\\
\underline{\textsf{Search}($w$)}\\[6pt]
\textcolor{blue}{Replacing lines 4-18 in \textsf{Search} in \textsf{SGX-SE1} with:}\\
1: \hspace*{9pt}\textcolor{blue}{\textbf{foreach} $id$ \textbf{in} $d$ \textbf{do}}\\
2: \hspace*{15pt}\textcolor{blue}{\textbf{if} ${\displaystyle BF[H^\prime_j(k_{BF},w \parallel id)]_{j \in [h]}=1}$}\\
3: \hspace*{22pt}\textcolor{blue}{$u^\prime\leftarrow H_3(k_w,id)$};\\
4: \hspace*{22pt}\textcolor{blue}{$v^\prime\leftarrow M_c[u^\prime]$};\\
5: \hspace*{22pt}\textcolor{blue}{$c\leftarrow \textsf{Dec}(k_{c},v^\prime)$};\\
6: \hspace*{22pt}\textcolor{blue}{$st_{w_c} \leftarrow \{c\} \cup st_{w_c}$};\\
7: \hspace*{22pt}\textcolor{blue}{delete $M_c[u^\prime]$};\\
8: \hspace*{22pt}\textcolor{blue}{delete $R[id]$}; // delete doc \\
%9: \hspace*{15pt}\textcolor{blue}{\textbf{end if}}\\
%10:\hspace*{9pt}\textcolor{blue}{\textbf{end foreach}} \\

\end{multicols}
\vspace{-30pt}
\end{framed}
\vspace{-10pt}
\caption{Protocols in \textsf{SGX-SE2}. The new instructions of \textsf{SGX-SE2} is in \textcolor{blue}{blue}}
\label{fig:X2_protocols}
\vspace{-15pt}
\end{figure*}

\begin{sloppypar}
In \textsf{setup}, client communicates with enclave upon an established secure channel to provision $K=(k_\Sigma,k_f)$ where $k_\Sigma$ enables enclave to generate update/query tokens and $k_f$ is the symmetric key for document encryption/decryption. The enclave maintains the maps $ST$ and $D$, and the list $d$, where $ST$ stores the states of keywords, $D$ presents the mapping between keywords and deleted documents, and $d$ is the array of deleted $id$s. The server holds an encrypted index $M_I$, the map of encrypted state $M_c$, and the repository $R$ with $R[id]$ stores the encrypted document of document identifier $id$.
\end{sloppypar}

In \textsf{update}, the client receives a tuple $(\textsf{op},\textsf{in})$, where it could be $(\textsf{op}=add, \textsf{in}=(\textsf{doc},id))$ or  $(\textsf{op}=del,\textsf{in}=id)$. If the update is addition, the client encrypts \textsf{doc} by using $k_f$ and sends that encrypted document to server. After that, the client sends $(\textsf{op},\textsf{in})$ to the enclave. The enclave will then parse \textsf{doc} to retrieve the list $L$ of $\{(w,id)\}$. For each $w$, the enclave generates $k_w$ and $k_c$ from $k_\Sigma$, and retrieves the latest state $c\leftarrow ST[w]$. The enclave will then generate $k_{id}$ from $c$ by using $H_1(k_w,c)$ with $H_1$ is a hash function. After that, the enclave uses $k_w$, $k_c$, and $k_{id}$ to generate encrypted entries $(u,v)$ and $(u^\prime,v^\prime)$ for $w$. In particular, the first encrypted entry, with $(u,v)\leftarrow (H_2(k_w,c),\textsf{Enc}(k_{id},id))$, holds the mapping between $c$ and $id$ to allows the server retrieves $id$ based on given $u$ and $k_{id}$. The second encrypted entry, with $(u^\prime,v^\prime)\leftarrow (H_3(k_w,id),\textsf{Enc}(k_{c},c))$, hides the state $c$ of documents. In this way, the client can retrieve the state $c$ of deleted documents upon sending $u^\prime$ in \textsf{search} operation. In our protocols, $H_1$ and $H_2$ are hash functions, and \textsf{Enc} is a symmetric encryption cipher. We note that enclave only sends a batch of $(T_1,T_2)$ to the server within one $ocall$ per a document addition, where $T1=\{(u_{w_1},v_{w_1}),\ldots,(u_{w_{|L|}},v_{w_{|L|}})\}$ and $T2=\{(u^\prime_{w_1},v^\prime_{w_1}),\ldots,(u^\prime_{w_{|L|}},v^\prime_{w_{|L|}})\}\}$. Then, the server will update $T_1$ and $T_2$ to $M_I$ and $M_c$, respectively. If the update is deletion, the enclave simply updates $d$ by the deleted $id$ without further computation or communication to the server.

In \textsf{search}, the client sends a query $q$ containing $w$ to the enclave via the secure channel and expects to receive all the \textit{current} (non-deleted) documents matching $w$ from the server. The enclave begins loading deleted encrypted documents in $d$ from the server in a sequential manner. By using $k_f$, the enclave decrypts those documents for checking the existence of $w$, and updating $D[w]$ if applicable. By leveraging $D[w]$, the enclave can retrieve the state list ${st_{w_c}}=\{c^{del}_{id}\}$, where $c^{del}_{id}$  is the state used when the enclave added the deleted document $id$ for $w$. After that, the enclave simply infers the states of non-deleted documents by excluding ${st_{w_c}}$ from the set of $\{0,\ldots,ST[w]\}$. Finally, the enclave will compute the query token $u$ and $k_{id}$ for these non-deleted documents, and send the list $Q_w=\{(u,k_{id})\}$ to the server. At the server, upon receiving $Q_w$, it can retrieve $id_i$ when decrypting $M_I[u_i]$ with $k_{id_i}$. Finally, the server returns the encrypted documents $\textit{Res}=\{R[id_i]\}$ to the client. 

\textbf{Efficiency of }\textsf{SGX-SE1:}
In \textsf{update}, \textsf{SGX-SE1} only takes $n_w$ \textit{ocalls} to add all $n$ documents containing $w$ to the server, and no \textit{ocall} for deletion due to the caching of deleted documents within the enclave. That efficiency outperforms \textsf{Bunker-B} since  the latter requires an additional $ocall$ per a deletion. However, we note that the asymptotic performance of \textsf{SGX-SE1} is affected by $(d+d_w)$ search roundtrips. In particular, the enclave needs to load and decrypt deleted documents within the enclave. Thus, the search performance really depends on how large the number of deleted documents is at the query time. We will later compare our search latency with \textsf{Bunker-B} in Section~\ref{sec:evaluation}.

%\subsection{\textsf{SGX-SE2}: Very Low Overhead with Efficient Searches}

\subsection{\textsf{SGX-SE2} Construction}\label{subsec:sgx2}

According to Table~\ref{tab:comparison}, \textsf{SGX-SE1} has $(d+d_w)$ search roundtrips and non-trivial $O(n_w +d)$ computation. One downside is that the enclave needs to spend time on decrypting deleted documents. Here, we present \textsf{SGX-SE2}, an advanced version of \textsf{SGX-SE1}, that reduces search roundtrips to $d_w$ and achieves better asymptotic and concrete search time $O(n_w + v_d)$. The main solution we make to \textsf{SGX-SE2} is that we use a Bloom filter \textit{BF} within the enclave to verify the mapping between query keyword $w$ and deleted document $id$s. In this way, \textsf{SGX-SE2} avoids loading them from the server. Since \textit{BF} is a probabilistic data structure, we can configure it to achieve a negligible false positive rate $P_e$ (see Section~\ref{sec:evaluation}). In Figure~\ref{fig:X2_protocols}, we highlight the solution of \textsf{SGX-SE2}. We summarily introduce \textsf{SGX-SE2} as follows:

In \textsf{setup}, \textsf{SGX-SE2} is almost the same with that one in \textsf{SGX-SE1} with the exception that the client also requires to initialise the parameters of \textit{BF}. They are, $k_{BF}$, $b$ and $h$, where $k_{BF}$ is the key for computing the hashed value of $(w||id)$, and $b$ is the number of bits in the \textit{BF} vector (i.e, vector size), and $h$ is the number of hash functions. Upon receiving the \textit{BF} setting, the enclave initialises the \textit{BF} vector and the set of hash functions $\{H_j^\prime\}_{j \in[h]}$. In \textsf{SGX-SE2}, the mapping $D$ between keywords and deleted $id$s is no longer needed within the enclave like that one in \textsf{SGX-SE1}.  

In \textsf{update}, \textsf{SGX-SE2} is also similar with \textsf{SGX-SE1}. However, if the update is addition, the enclave computes a new member $H^\prime_j(k_{BF},w \parallel id)$ to update \textit{BF}.

In \textsf{search}, \textsf{SGX-SE2} verifies the mapping between query keyword $w$ and deleted \textit{id}s by checking the membership of $(w||id)$ with \textit{BF}. If the mapping is valid, \textsf{SGX-SE2} performs the same as \textsf{SGX-SE1} to retrieve the state list ${st_{w_c}}=\{c^{del}_{id}\}$, where $c^{del}_{id}$  is the state used for deleted $id$s. After that, the enclave infers the states of non-deleted documents and computes query tokens to send to the server.

\noindent \textbf{Efficiency of} \textsf{SGX-SE2:} The scheme clearly outperforms \textsf{SGX-SE1} in terms of search computation and communication roundtrips due to the usage of the Bloom filter. It avoids loading $d$ deleted documents into the enclave, making the search roundtrip only $d_w$. The scheme is even more efficient when $|d|$ is large. The reason is that the cost of verifying a membership $(w||id)$ is always $O(1)$ under the fixed \textit{BF} setting. We note that checking $d$ members in the \textit{BF} is still more efficient than loading/decrypting their real documents. \textit{BF} is also memory-efficiently; therefore, one can configure its size to balance the enclave memory with the demand of large datasets.

\noindent\textbf{Remark}: 
%
%Note that \textsf{Bunker-B} takes extra $d$ \textit{ocall}s in non-batch processing to physically delete  $d$ documents, while \textsf{SGX-SE1} and \textsf{SGX-SE2} take the \textit{ocall}.
%
%But, upon receiving these \textit{ocall} instructions, the server cost to physically delete the documents in the document repository $R$ should be same across the three schemes. 
%
%This physical cost just depends on the server power. 
%
%Note that, the three schemes can also perform deleting multiple documents in a batch.
%
Note that deleting a document \textsf{doc} with identifier $id$ in \textsf{Bunker-B} requires deletion entries of all keywords in that \textsf{doc} with ($w_i,id, \textsf{op}=del$) have been inserted in the encrypted index $M_I$ beforehand. That would require $M$ $ocall$s for the \textsf{doc} of $M$ keywords.
Then, \textsf{Bunker-B} takes extra one \textit{ocall}  to physically delete the \textsf{doc}. 
This physical deletion cost is the same with \textsf{SGX-SE1} and \textsf{SGX-SE2} (i.e., one $ocall$) except that these two schemes do not require any deletion entries to be inserted in $M_I$.
Clearly, \textsf{Bunker-B}, \textsf{SGX-SE1}, and \textsf{SGX-SE2} can do batch processing to delete $d$ documents in one $ocall$.
With \textsf{SGX-SE1}, deleting a \textsf{doc} can be done right after all keywords in the deleted document have been cached in $D[w]$ (see Figure~\ref{fig:X1_protocols}). 
With \textsf{SGX-SE2}, a \textsf{doc} can be deleted at the earliest time when any keyword in the \textsf{doc} is being searched (see Figure~\ref{fig:X2_protocols}).

\section{Security Analysis}
\label{sec:security}

\textsf{SGX-SE1} and \textsf{SGX-SE2} contain the leakage of search and updates, because the server can observe the interaction between its memory and the enclave. In \textsf{setup}, the schemes leak nothing due to the remote attestation and secure data communication channel between the client and the enclave. If  \textsf{update} is addition, the server is able to track the time and memory access when new entries are inserted into data structures $M_I$, $M_c$, and $R$ (see Fig.~\ref{fig:X1_protocols} for their definitions). If  \textsf{update} is deletion, the enclave does not communicate with the server during \textsf{deletion}. Hence, there is no leakage in the operation. In \textsf{search}, the access patterns on $M_I$, $M_c$, and $R$ are revealed to the server. 
The only difference between \textsf{SGX-SE1} and \textsf{SGX-SE2} in term of security is that \textsf{SGX-SE1} requires to load encrypted deleted documents to the enclave during the search. Therefore, our following analysis is almost identical to both schemes. We will state the difference between them wherever is necessary. 

We formulate the detail leakage and define the $\textbf{Real}_{\mathcal{A}}(\lambda)$ and a $\textbf{Ideal}_{\mathcal{A},\mathcal{S}}(\lambda)$ game for an adaptive adversary $\mathcal{A}$ and a polynomial time simulator $\mathcal{S}$ with the security parameter $\lambda$ as follows. 

\begin{sloppypar}
	We denote $\mathcal{D}$ as our general scheme that could be \textsf{SGX-SE1} or \textsf{SGX-SE2}. The security of $\mathcal{D}$  can be quantified via a stateful leakage function $\mathcal{L}=(\mathcal{L}^{Stp},\mathcal{L}^{Updt},\mathcal{L}^{Srch},\mathcal{L}^{hw})$. 
	The first three components define the information exposed in \textsf{Setup}, \textsf{Update}, and \textsf{Search}, respectively.
	The latter one, $\mathcal{L}^{hw}$, defines the inherent leakage of the used SGX enclave with the outputs from the enclave to the server. 
	We now define $\mathcal{L}$ and then formalise our security with analysis.
\end{sloppypar}

In \textsf{Setup}, $\mathcal{D}$ leaks nothing to the server except the data structure of $M_I$ (i.e., the encrypted index), $M_c$ (i.e., the encrypted map of keyword states), $R$ (i.e., the empty repository of encrypted documents).
%
%In \textsf{Update}, if $\textsf{op}=add$, $\mathcal{D}$ is \textit{forward-private} as presented in %Figure.~\ref{fig:X1_protocols}. Hence, $\mathcal{L}^{Updt}$ can be quantified as \[
%		\mathcal{L}^{Updt}(\{(\textsf{op},\textsf{in})\}) = \mathcal{L}^\prime(\{(id_i,\mu_i)\})
%\]
%\noindent where $\{(id_i,\mu_i)\}$ captures the number of entries to be added into $M_I$ and $M_c$; and %$\mathcal{L}^\prime$ is stateless.

In \textsf{Update}($\textsf{op}=add,\textsf{in}$), $\mathcal{D}$ leaks the data access pattern of encrypted entries to be inserted in $M_I$, $M_c$, and $R$. Otherwise, if $\textsf{op}=del$, $\mathcal{D}$ leaks nothing under the secure channel established in \textsf{Setup}. Hence,
\[
    \mathcal{L}^{Updt}(\{(\textsf{op},\textsf{in})\})=\{(T_1,T_2,R[id_i])\}
\]
where $T_1=\{(u,v)\}$ and $T_2=\{(u^\prime,v^\prime)\}$ present the collections of entries to be inserted in $M_I$ and $M_c$ respectively, and $R[id_i]$ denotes an encrypted document to be inserted in $R$ with label $id_i$.

In \textsf{Search}($w$), $\mathcal{D}$ leaks 1) the access pattern on $M_c$ when the enclave queries the deleted states of $w$, named $\textsf{ap}_{M_c}(w)$, 2) the access pattern on $M_c$ when the enclave queries non-deleted $id$s, named $\textsf{ap}_{M_I}(w)$, if $\mathcal{D}$ is \textsf{SGX-SE1}, and 3) the pattern on deleted documents $d_w$, named $\textsf{ap}_R(d_w)$. Then, formally
\[		
		\mathcal{L}^{Srch}(w) = \textsf{ap}_{M_c}(w) + \textsf{ap}_{M_I}(w) + [\textsf{ap}_R(d_w)]
\]
We define $\mathcal{L}^{hw}(M_I,M_c,R)$ as the hardware leakage during \textsf{Update} and \textsf{Search}. That includes memory access and location, the time log, and the size of the manipulated memory area.
\[
		\mathcal{L}^{hw}(M_I,M_c,R) = (M_I,M_c,R)^\textit{Updt} + (M_I,M_c,R)^\textit{Srch}
\]
This function outputs the trace $\tau$ of $(l,T,v,t)$, where $l$ is the label input, $T$ is a map data structure that could be $M_I$, $M_c$, and $R$, $v$ is the value at $T[l]$, and $t$ is the time access of \textsf{op}. W.r.t. \textsf{SGX-SE1}, if $l$ is an $id$, the function will output the encrypted document $e$ and the document size $|e|$.

\begin{definition} 
\label{def:security}
\begin{sloppypar}
	Let $\mathcal{D}$ denote our scheme that consists of three protocols \textsf{Setup}, \textsf{Update}, and \textsf{Search}. Consider the probabilistic experiments $\textbf{Real}_{\mathcal{A}}(\lambda)$ and $\textbf{Ideal}_{\mathcal{A},\mathcal{S}}(\lambda)$, whereas $\mathcal{A}$ is a stateful adversary, and $\mathcal{S}$ is a stateful simulator that gets the leakage function $\mathcal{L}$.
\end{sloppypar}

$\textbf{Real}_{\mathcal{A}}(\lambda)$: The challenger runs $\textsf{Setup}(1^\lambda)$ that involves the client, the enclave, and the server to initialise necessary data structures as presented in Figure.~\ref{fig:X1_protocols}. $\mathcal{A}$ chooses a database 
$\textsf{DB}=\{\textsf{doc}_i\}_{i\in Z}$ and makes a polynomial number of updates (addition/deletion) with $(\textsf{op},\textsf{in})$, where $Z$ is a natural number of documents, and $(\textsf{op}=add,in=\textsf{doc}_i)$ or $(\textsf{op}=del,\textsf{in}=id_i)$. Accordingly, the challenger runs those updates with $\textsf{Update}(\textsf{op},\textsf{in})$ and eventually returns the tuple $(M_I,M_c,R)^\textit{Updt}$ to $\mathcal{A}$. After that, $\mathcal{A}$ adaptively chooses the keyword $w$  (\textit{resp.}, $(\textsf{op},\textsf{in})$) to search (\textit{resp.}, update). In response, the challenger runs \textsf{Search}($w$) (\textit{resp.}, \textsf{Update}(\textsf{op},\textsf{in})) and returns the transcript of each operation. The challenger also returns $(M_I,M_c,R)^\textit{Srch}$ to $\mathcal{A}$. Finally, $\mathcal{A}$ outputs a bit $b$.

\begin{sloppypar}
	$\textbf{Ideal}_{\mathcal{A},\mathcal{S}}(\lambda)$: $\mathcal{A}$ chooses a $\textsf{DB}=\{\textsf{doc}_i\}_{i\in Z}$. By using $\mathcal{L}^{Updt}$ and $(M_I,M_c,R)^\textit{Updt} $, $\mathcal{S}$ creates a tuple of $(M_I,M_c,R)$ and passes it to $\mathcal{A}$. Then, $\mathcal{A}$ adaptively chooses the keyword $w$  (\textit{resp.}, $(\textsf{op},\textsf{in})$) to search (\textit{resp.}, update). The challenger returns the transcript simulated by $\mathcal{S}(\mathcal{L}^{Srch}(w))$ (\textit{resp.}, $\mathcal{S}(\mathcal{L}^{Updt}(\textsf{op},\textsf{in}))$) with $(M_I,M_c,R)^\textit{Srch}$. Finally, $\mathcal{A}$ returns a bit $b$.
\end{sloppypar}

We say $\mathcal{D}$ is $\mathcal{L}$-secure against adaptive chosen-keyword attacks if for all probabilistic polynomial-time algorithms $\mathcal{A}$, there exist a PPT simulator $\mathcal{S}$ such that
\[
		|Pr[\textbf{Real}_{\mathcal{A}}(\lambda) = 1] - Pr[\textbf{Ideal}_{\mathcal{A},\mathcal{S}}(\lambda) = 1]| \leq negl(\lambda)
\]

\end{definition}  

\begin{theorem}\label{theo:theorem_sgx}
The scheme $\mathcal{D}$ presented above is $\mathcal{L}$-secure according to Def~\ref{def:security}.
\end{theorem}

We proof the schemes are secure if they achieve both forward privacy and Type-II backward privacy. We note that the client issues a query on $w$ to the enclave via an established secure channel. Hence, $\mathcal{A}$ has to generate a query token by itself in the game of Def.~\ref{def:security}. Regarding forward privacy, the increasing   state $ST[w]$ (see Fig.~\ref{fig:X1_protocols}) when adding a new document containing $w$ ensures that $\mathcal{A}$ cannot generate a query token to retrieve a newly added document. 
W.r.t. backward privacy, $\mathcal{A}$ statistically knows the timestamps when the deleted states of $w$ added in $M_c$ when the enclave requests the server to access to $M_c$ during \textsf{search}. However,  $\mathcal{A}$  does not know when they were requested for \textsf{deletion} by the client. The reason for that is because \textsf{SGX-SE1} and \textsf{SGX-SE2} cache these deletion requests in the enclave and only access them during \textsf{search}. As a result, $\mathcal{A}$  does not know which delete updates occur and have cancelled addition updates.

We now prove Theorem~\ref{theo:theorem_sgx} by describing a PPT simulator $\mathcal{S}$ for which a PPT adversary $\mathcal{A}$ can distinguish $\textbf{Real}_{\mathcal{A}}(\lambda)$ and $\textbf{Ideal}_{\mathcal{A},\mathcal{S}}(\lambda)$ with negligible probability.
\begin{proof} $\mathcal{S}$ first generates a random key $\tilde{K}=(\tilde{k}_\Sigma,\tilde{k}_f)$ to simulate the key components that the enclave contains (see Figure~\ref{fig:X1_protocols}). Then, $\mathcal{A}$ executes \textsf{Search}(w) with $w$, which is a random keyword, in order to obtain a query token $q$ sent by the enclave. Then, $\mathcal{A}$ simulates addition tokens $a$ for $w$ based on $\tilde{K}$ and $\mathcal{L}^{hw}(M_I,M_c,R)$, and sends them to the enclave to receive the new update of $(M_I,M_c,R)$. However, $\mathcal{A}$ cannot map which update token in $a$ relates to $q$. The reason is that the enclave keeps increasing the state $ST[w]$. Hence, $\mathcal{A}$ cannot distinguish between the output of $\textbf{Real}_{\mathcal{A}}(\lambda)$ and the simulated output in \textsf{Update} and \textsf{Search} (\textit{forward privacy}). 

During \textsf{Search}, if there were delete updates made in the past on deleted documents $d$ with identifier list $\{id_i\}$,  $\mathcal{A}$ cannot know which keywords are inside the encrypted doc $R[id_i]$. Also, $\mathcal{A}$ does not know when delete updates made since the enclave only requests $d$ during \textsf{Search}. The $\textsf{ap}_{M_c}(w)$ does not reveal $id_i$ (see \textsf{Search} in Fig~\ref{fig:X1_protocols}). However, $\mathcal{A}$ knows the time when the entry relating $id_i$ added to ${ap}_{M_c}$ via $\mathcal{L}^{hw}$, and how many $id_i$ in $d$. Clearly, at the end of the protocol $\mathcal{A}$ knows how many current (non-deleted) $id$ accessed. Hence, $\mathcal{D}$ is type-II \textit{backward privacy}.
%deletion) tokens based on $\tilde{K}$ and $\mathcal{L}^{hw}(M_I,M_c,R)$ and sends them to the enclave to receive the new update of $(M_I,M_c,R)$. After that, $\mathcal{S}$ generates random query tokens and sends them to the enclave via \textsf{Search} protocol of $\mathcal{D}$ to receive $(M_I,M_c,R)^\textit{Srch}$ and the transcripts of the queries. However, $\mathcal{A}$ cannot distinguish between the output of $\textbf{Real}_{\mathcal{A}}(\lambda)$ and the simulated output in \textsf{update} and \textsf{search} since $\mathcal{D}$ maintains \textit{forward} and Type-II \textit{backward privacy}.
\end{proof}

\section{Implementation and Evaluation}
\label{sec:evaluation}
%\vspace{-5pt}
\noindent \textbf{Experiment setup and implementation}: We choose two datasets: One is a synthesis dataset ($3.2$ GB) generated from the English keyword frequency data based on the Zipf's law distribution, and the other one is the Enron email dataset\footnote{Enron email dataset: \url{https://www.cs.cmu.edu/~./enron/}} ($1.4$ GB).
A summary of the datasets is given in Table~\ref{tlb:statistics}.

We build the prototype of \textsf{SGX-SE1} and \textsf{SGX-SE2} using C++ and the Intel SGX SDK\footnote{Source code: \url{https://github.com/MonashCybersecurityLab/SGXSSE}}.
In addition, we implement the prototype of \textsf{Bunker-B} as the baseline for comparisons, since its implementation is not publicly available.
The prototype leverages the built-in cryptographic primitives in the SGX SDK to support the required cryptographic operations. 
It also uses the settings and APIs from the SDK to create, manage and access the application (enclave) designed for SGX.
Recall that the SGX can only handle $96$ MB memory within the enclave. 
Access to the extra memory space triggers the paging mechanism of the SGX, which brings an extra cost to the system.% (average $5\times$ as reported in~\cite{taassori2018vault}).
To avoid paging in our prototype, our prototypes are implemented with batch processing to tackle with the keyword-document pairs, which splits a huge memory demand into multiple batches with smaller resource requests.
The batch processing enables our prototypes to handle queries with large memory demands.
Moreover, the prototype should avoid too many \textit{ecall}s/\textit{ocall}s as it incurs the I/O communication cost between the untrusted and the trusted application (enclave).
Hence, in the following experiments, we set the batch size to $1\times 10^5$ for all schemes, which can avoid triggering paging while minimising the number of \textit{ecall}/\textit{ocall}s.
The prototypes are deployed in a workstation equipped SGX-enabled Intel i7 2.6 GHz and 32 GB RAM. 
\vspace{-5pt}

\begin{table}[!t]
	\caption{Statistics of the datasets used in the evaluation.}
	\label{tlb:statistics}
	\centering
	\begin{tabular}{|c|c|c|c|}
		\hline        		
		Name & \# of keywords & \# of docs & \# of keyword-doc pairs \\		
		\hline
		Synthesis & $1,000$ & $1,000,000$ & $11,879,100$ \\
		\hline
		Enron & $29,627$ & $517,401$ & $37,219,800$ \\
		\hline
 	\end{tabular}
\end{table}

\begin{table}[!t]
	\caption{Avg. ($\mu$s) for adding a keyword-doc pair under different schemes.}
	\label{tlb:setup}
	\centering
	\begin{threeparttable}[!t]
	\begin{tabular}{|c|c|c|c|c|}
		\hline        		
		\# of docs & \# of keyword-doc pairs & BunkerB & SGX-SE1 & SGX-SE2 \\
		\hline
		$2.5\times 10^5$ & $2.5\times 10^5$ & $21$ & $23$ & $26$ \\
		\hline
		$5\times 10^5$ & $6.5\times 10^5$ & $19$ & $19$ & $21$ \\
		\hline
		$7.5\times 10^5$ & $1.9\times 10^6$ & $15$ & $12$ & $14$ \\
		\hline
		$1\times 10^6$ & $1.18\times 10^7$ & $12$ & $7$ & $8$ \\
		\hline
 	\end{tabular}
 	 \begin{tablenotes}
		\item $\star$: The average time decreases since the average I/O cost of loading keywords from the file decreases
	\end{tablenotes}
	\end{threeparttable}
\end{table}

\subsection{Performance evaluation on the synthesis dataset}\label{subsec:synthesis}
\noindent \textbf{Insertion and deletion}: First, we evaluate the time for insertion and deletion under three different schemes.
We follow a reversed Zipf's law distribution to generate the encrypted database of our synthesis dataset, and we measure the runtime for adding one keyword-document pair into the encrypted database of different schemes.
As shown in Table~\ref{tlb:setup}, \textsf{Bunker-B} takes $21\ \mu$s to insert one pair, which is faster than our schemes ($23\ \mu$s and $26\ \mu$s) when the number of keyword-document pairs equals the number of documents.
The reason is that the insertion time of the above three schemes is bounded by the I/O (\textit{ecall}/\textit{ocall}) between the untrusted application and the enclave.
For \textsf{Bunker-B}, the I/O cost is linear to the number of keyword-document pairs (see Sec.\ref{subsec:design} for details), while the one for our schemes is linear to the number of documents.
Also, our schemes involve more computations (PRF, Hash) and maintain more data structures (Bloom filter), which require more time to be processed.
Nonetheless, when inserting $1\times 10^6$ documents, our schemes only require $7\ \mu$s and $8\ \mu$s respectively to insert one keyword-document pair, which is $2\times$ faster than \textsf{Bunker-B} ($12\ \mu$s).
In the above case, the number of keyword-document pairs is $10\times$ larger than the number of documents, which implies that \textsf{Bunker-B} needs $10\times$ more I/O operations (\textit{ecall}/\textit{ocall}) to insert the whole dataset comparing to our schemes (see Table~\ref{tlb:insert_io} for details).
Note that the real-world document typically consists of more than one keyword. 
Hence, our schemes are more efficient than \textsf{Bunker-B} when dealing with a real-world dataset (see Section~\ref{subsec:enron}).

For deletion, the performance of \textsf{Bunker-B} is identical to that for insertion ($12\ \mu$s), because deletion runs the same algorithm with different operations.
For our schemes, the deletion process only inserts the document \textit{id} into a list, and the deletion operation is executed by excluding the deleted \textit{id} during the query phase.
Thus, our schemes only need $4\ \mu$s to process one doc in deletion phase.

\begin{figure}[!t]
	\centering
	\subfloat[25\% deletion] {
	\includegraphics[width=0.61\linewidth]{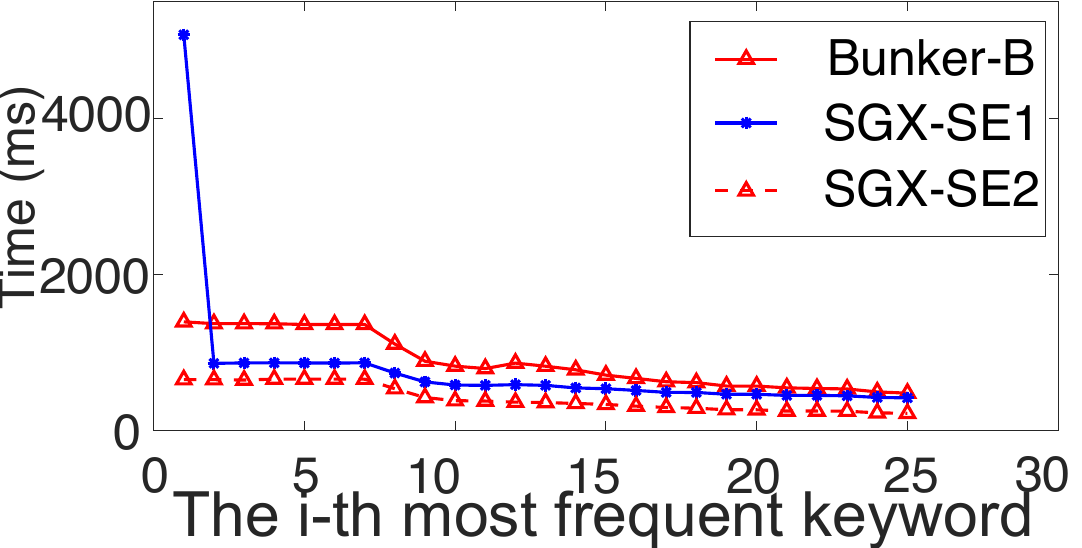}
		\label{fig:250k_25}
	}
	\hfil
	\subfloat[50\% deletion] {
	\includegraphics[width=0.61\linewidth]{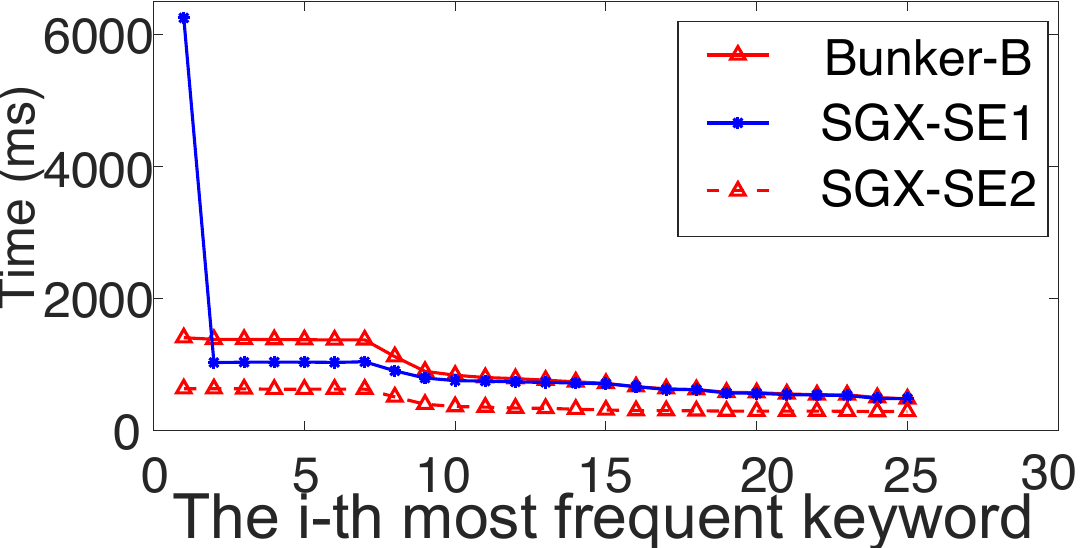}
		\label{fig:250k_50}
	}
	\hfil
	\subfloat[75\% deletion] {
	\includegraphics[width=0.61\linewidth]{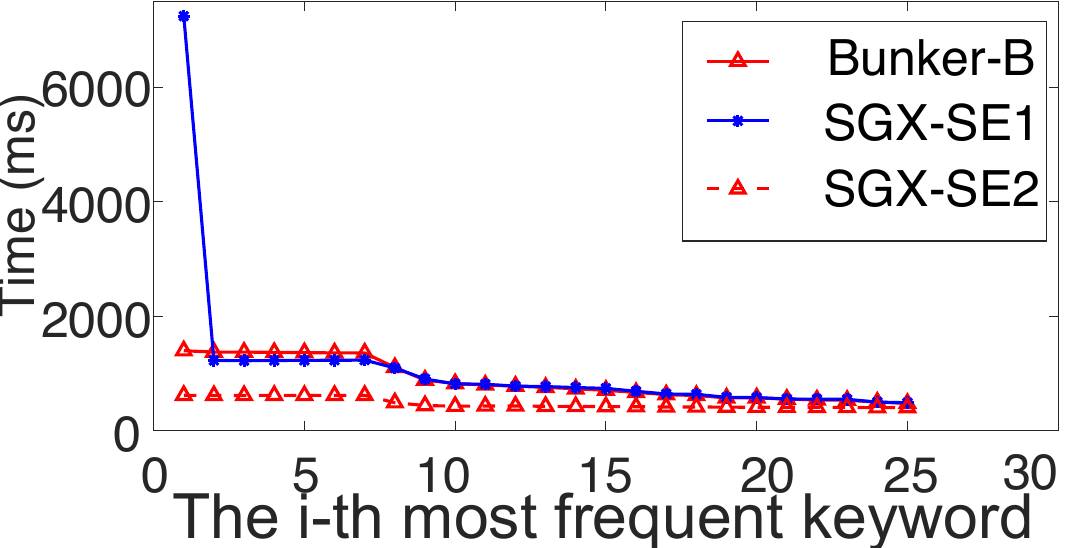}
		\label{fig:250k_75}
	}
	\caption{The query delay of querying the i-th most frequent keyword in the synthesis dataset under different schemes (insert $2.5\times 10^5$ docs and delete a portion).} \label{fig:query_250k}
	
\end{figure}

\noindent \textbf{Query delay}: Next, we report the query delay comparison between \textsf{Bunker-B} and our schemes to show the advantage of using \textsf{SGX-SE1} and \textsf{SGX-SE2}.
To measure the query delay introduced by keyword frequency and the deletion operation, we choose to query the top-25 keywords after deleting a portion of documents.
In our first evaluation, we insert $2.5\times 10^5$ documents and delete 25\%, 50\% and 75\% of the documents, respectively.
Fig.~\ref{fig:250k_25} illustrates the query delays when deleting 25\% of documents:
For the most frequent keyword, \textsf{Bunker-B} needs $1.3$ s to query while \textsf{SGX-SE2} only needs $654$ ms.
Although \textsf{SGX-SE1} takes $5$ s to perform the first search, it also caches the deleted keyword-document pairs inside the enclave and performs deletion on documents during the first query.
As a result, the rest of the queries are much faster, as the number of \textit{ocall}s is significantly reduced ($900\ \mu$s if we query the most frequent keyword again).
Even for the 25-th most frequent keyword, \textsf{SGX-SE1} ($159$ ms) and  \textsf{SGX-SE2} ($155$ ms) are still $40$\% faster than \textsf{Bunker-B} ($221$ ms).
\textsf{Bunker-B} is always slower than \textsf{SGX-SE1} and \textsf{SGX-SE2} in the above case as it requires to re-encrypt the remaining $75$\% documents after each query.
Compared to \textsf{Bunker-B}, \textsf{SGX-SE1} and \textsf{SGX-SE2} only access the deleted $25$\% files and exclude the corresponding token of deleted files before sending the token list (see Section~\ref{subsec:sgx2}).
With the increase of the deletion portion, the difference of the query delay between our schemes and \textsf{Bunker-B} becomes smaller as \textsf{Bunker-B} has fewer documents to be re-encrypted after queries.
When 75\% of the documents are deleted, our schemes still outperform \textsf{Bunker-B} when querying the keywords with a higher occurrence rate (see Fig.~\ref{fig:250k_75}).
However, their performances are almost the same when querying the 25-th most frequent keyword, i.e., about $400$ ms for three schemes, because \textsf{Bunker-B} only re-encrypts a tiny amount of document id (almost $0$).

\begin{figure*}[!t]
	\centering
	\subfloat[25\% deletion] {
	\includegraphics[width=0.31\linewidth]{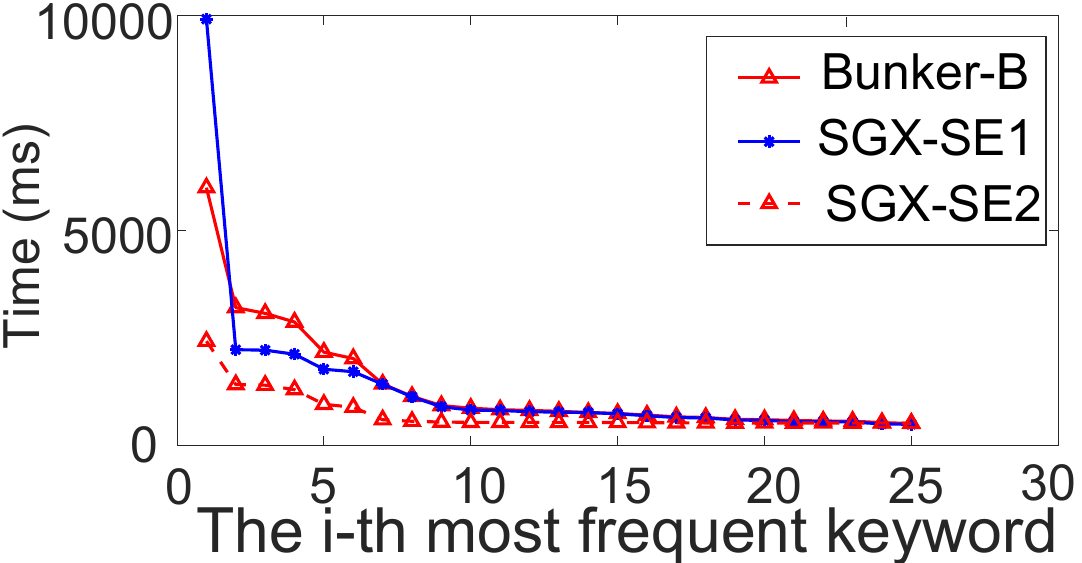}
		\label{fig:1m_25}
	}
	\hfil
	\subfloat[50\% deletion] {
	\includegraphics[width=0.31\linewidth]{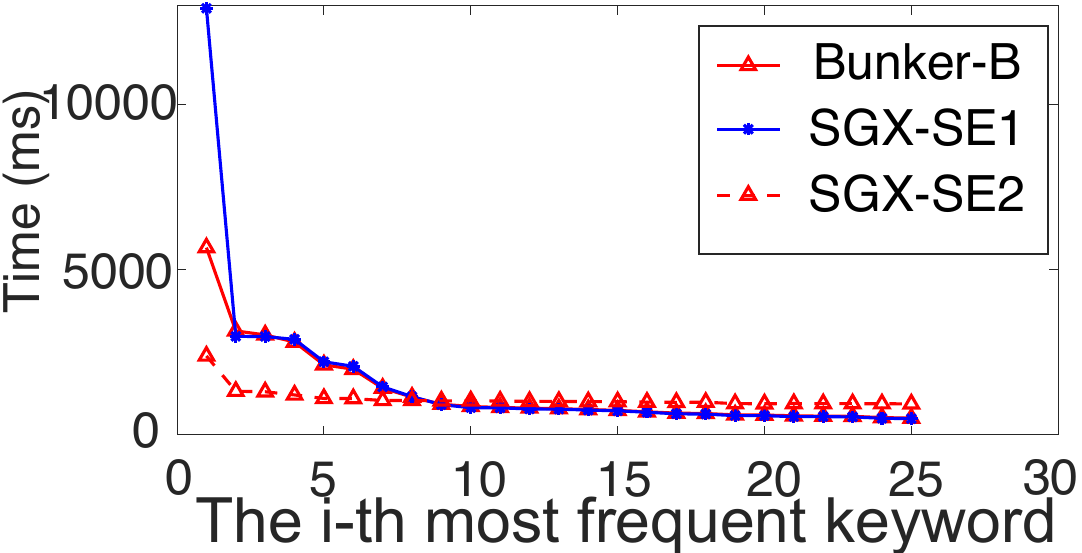}
		\label{fig:1m_50}
	}
	\hfil
	\subfloat[75\% deletion] {
	\includegraphics[width=0.31\linewidth]{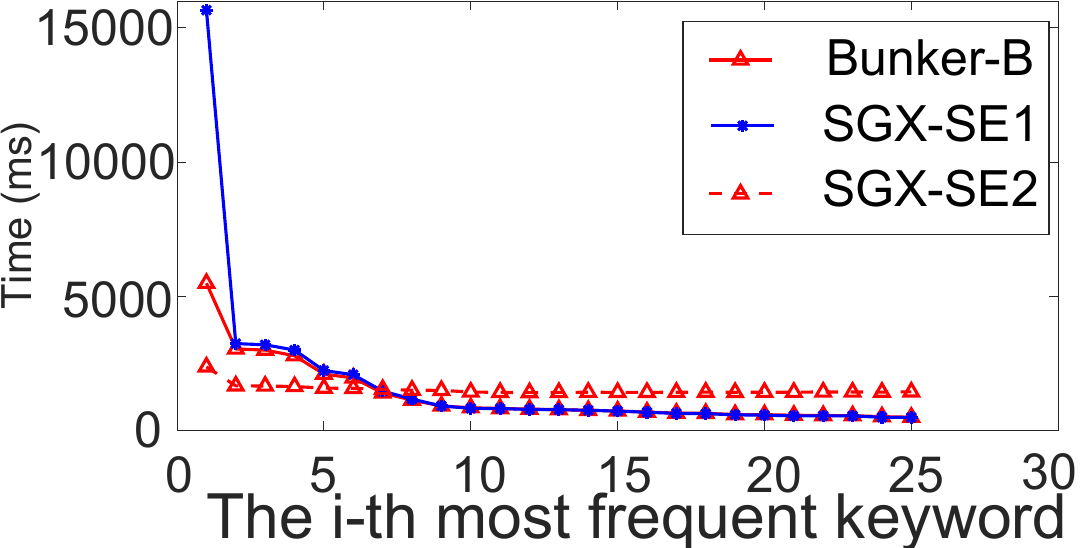}
		\label{fig:1m_75}
	}
	\caption{The query delay of querying the i-th most frequent keyword in the synthesis dataset (insert $1\times 10^6$ documents and delete a portion of them).} \label{fig:query_1m}
	\vspace{-15pt}
\end{figure*}

The second evaluation shows the query delay when inserting all $1\times 10^6$ documents into the encrypted database.
The major difference between this experiment and the previous one is that \textsf{SGX-SE1} scheme requires more than $128$ MB to cache the deleted documents, which triggers paging.
As shown in Fig.~\ref{fig:1m_25}, \textsf{SGX-SE1} needs $10$ s to cache the deleted documents.
When processing the query that contains a large number of documents (e.g., the second most frequent keyword), \textsf{SGX-SE1} ($2.4$ s) is almost $2\times$ slower than \textsf{SGX-SE2} ($1.4$ s).
Nonetheless, their query performance is still better than \textsf{Bunker-B}, which takes $3.2$ s to answer the above query.
When our schemes delete a larger portion of documents (see Figs~\ref{fig:1m_50} and~\ref{fig:1m_75}), the query delay of \textsf{SGX-SE1} and \textsf{SGX-SE2} is very close, since \textsf{SGX-SE1} only refers to the small deletion information cached in the enclave while \textsf{SGX-SE2} requires to check the Bloom filter for each deleted document.

\begin{table}[!t]
	\caption{Number of \textit{ecall}/\textit{ocall} for adding $1\times 10^6$ documents for different schemes.}
	\centering
	\label{tlb:insert_io}
	\begin{tabular}{|c|c|c|c|}
		\hline        		
		\# of calls & BunkerB & SGX1 & SGX2 \\
		\hline
		\textit{ecall} & $1.18\times 10^7$ & $1\times 10^6$ & $1\times 10^6$ \\
		\hline
		\textit{ocall} & $1.18\times 10^7$ & $1\times 10^6$ & $1\times 10^6$ \\
		\hline
 	\end{tabular}
 	%\vspace{-15pt}
\end{table}

\begin{table}[!t]
	\caption{Number of \textit{ecall}/\textit{ocall} for deleting a portion of documents after adding $1\times 10^6$ documents.}
	\centering
	\label{tlb:del_io}
	\tabcolsep 0.02in
	\begin{tabular}{|c|c|c|c|c|c|c|}
		\hline        		
		\multirow{2}{*}{Deletion \%} & \multicolumn{2}{c}{BunkerB} & \multicolumn{2}{|c|}{SGX1} & \multicolumn{2}{c|}{SGX2} \\
		\cline{2-7}   
		& \textit{ecall} & \textit{ocall} & \textit{ecall} & \textit{ocall} & \textit{ecall} & \textit{ocall} \\
		\hline
		25\% & $9.9\times 10^6$ & $9.9\times 10^6$ & $2.5\times 10^5$ & $0$ & $2.5\times 10^5$ & $0$\\
		\hline
		50\% & $1.12\times 10^7$ & $1.1\times 10^7$ & $5\times 10^5$ & $0$ & $5\times 10^5$ & $0$ \\
		\hline
		75\% & $1.16\times 10^7$ & $1.16\times 10^7$ & $7.5\times 10^5$ & $0$ & $7.5\times 10^5$ & $0$\\
		\hline
 	\end{tabular}
 	\vspace{-15pt}
\end{table}

\begin{table}[!t]
	\centering
	\caption{Number of \textit{ecall}/\textit{ocall} when querying the most frequent keyword after adding $1\times 10^6$ documents and deleting a portion of them.}
	\label{tlb:query_io}
	\begin{threeparttable}[!t]
	\begin{tabular}{|c|c|c|c|c|c|c|}
		\hline        		
		\multirow{2}{*}{Deletion \%} & \multicolumn{2}{c}{BunkerB} & \multicolumn{2}{|c|}{SGX1} & \multicolumn{2}{c|}{SGX2} \\
		\cline{2-7}   
		& \textit{ecall} & \textit{ocall} & \textit{ecall} & \textit{ocall} & \textit{ecall} & \textit{ocall} \\
		\hline
		25\% & $1$ & $21$ & $1$ & $250,011$$^\star$/$11$ & $1$ & $11$ \\
		\hline
		50\% & $1$ & $20$ & $1$ & $500,010$$^\star$/$10$ & $1$ & $10$ \\
		\hline
		75\% & $1$ & $21$ & $1$ & $750,011$$^\star$/$11$ & $1$ & $11$ \\
		\hline
 	\end{tabular}
 	\begin{tablenotes}
		\item $^\star$: It includes the \textit{ocall} for caching and deleting the encrypted documents.
	\end{tablenotes}
	\end{threeparttable}
	\vspace{-15pt}
\end{table}

%\vspace{-25pt}
\noindent \textbf{Communication cost}: The next evaluation demonstrates the impact of I/O operation (\textit{ecall}/\textit{ocall}) on the performance of different schemes.
As shown in Table~\ref{tlb:insert_io}, \textsf{Bunker-B} needs $10\times$ more \textit{ecall}/\textit{ocall} operations than our schemes.
Consequently, although both \textsf{Bunker-B} and our schemes generate and store the encrypted keyword-document pairs at the end, our schemes can achieve a better performance for insertion, because our schemes rely on less I/O operations.
This result is consistent with the average insertion time reported in the insertion and deletion part.
In terms of the deletion operation, \textsf{Bunker-B} needs almost $30\times$ more I/O operation than ours (see Table~\ref{tlb:del_io}).
Moreover, the deletion in our schemes only requires to insert the deleted id, which does not involve any cryptographic operation, whereas \textsf{Bunker-B} executes the same procedure as insertion.
This indicates that our schemes also have less communication cost than \textsf{Bunker-B}.
We further present the number of \textit{ecall}/\textit{ocall} involved during the query process in Table~\ref{tlb:query_io}.
Note that we implement batch processing for all schemes, so each \textit{ocall} can process $10^5$ query tokens at the same time.
The result shows that \textsf{Bunker-B} has more \textit{ocall} during the query process because it needs to issue tokens to query all document id as well as the deleted document.
After that, it should issue additional tokens to re-encrypt the undeleted documents.
On the other hand, our schemes keep the state map within the enclave, which indicates that our schemes do not require to retrieve all the document id via \textit{ocall}.
In most of the case, \textsf{Bunker-B} has $2\times$ more I/O operations than our schemes except for the cache stage of \textsf{SGX-SE1}.
Despite the fact that \textsf{SGX-SE1} takes more than $10^5$ \textit{ocall}s to perform caching, we stress that this is a one-time cost; it also enables our scheme to remove the document physically, whereas \textsf{Bunker-B} only can delete the document from the encrypted index.
%one-time amortised for search
%bosu our scheme the ocall already include deleted real files.

\begin{figure*}[!t]
	\centering

	\subfloat[25\% deletion] {
	\includegraphics[width=0.31\linewidth]{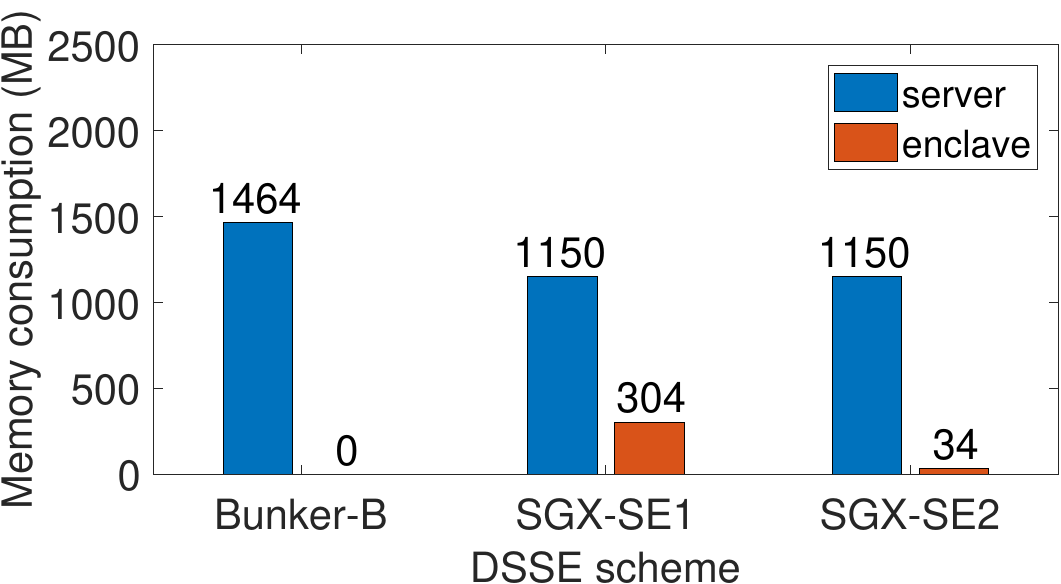}
		\label{fig:memory_1m_25}
	}
	\hfil
	\subfloat[50\% deletion] {
	\includegraphics[width=0.31\linewidth]{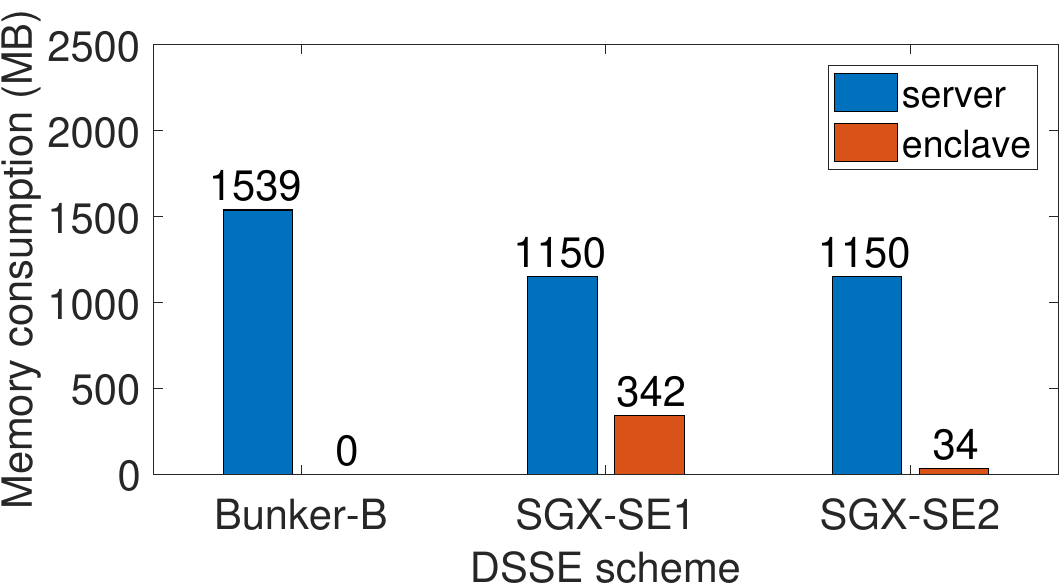}
		\label{fig:memory_1m_50}
	}
	\hfil
	\subfloat[75\% deletion] {
	\includegraphics[width=0.31\linewidth]{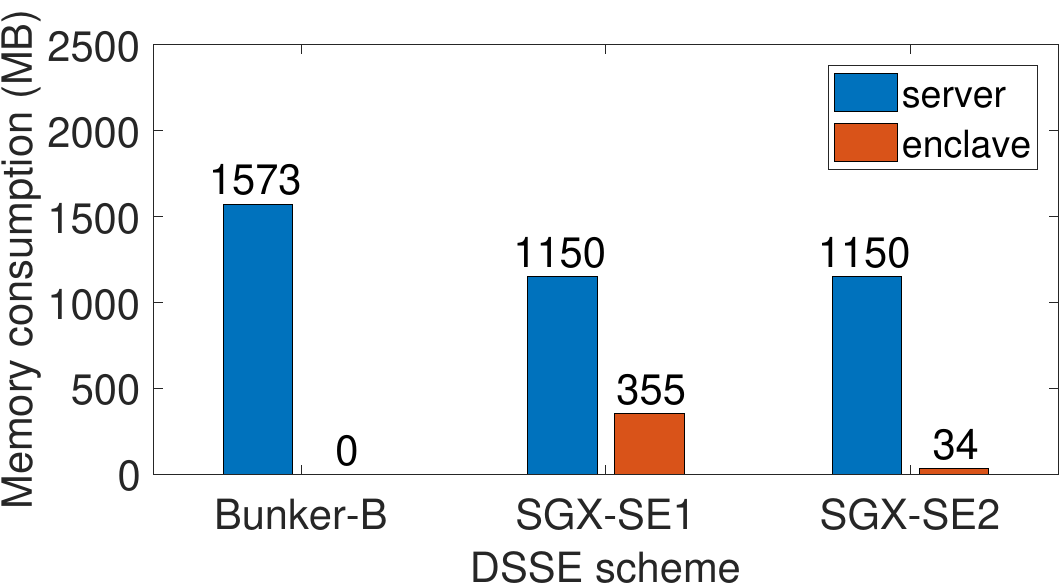}
		\label{fig:memory_1m_75}
	}
	
	\caption{The memory consumption in the synthesis dataset (inserted $1\times 10^6$ documents and deleted a portion of them).}\label{fig:memory}
	\vspace{-20pt}
\end{figure*}

\noindent \textbf{Memory consumption}: Finally, we present the memory consumption of three different schemes.
Since the memory consumption on the client is negligible comparing to  that for the server and enclave (i.e., less than $1$ MB).
As shown in Fig.~\ref{fig:memory}, the encrypted database always keeps unchanged for \textsf{SGX-SE1} and \textsf{SGX-SE2} because they keep the same keyword-document pairs after adding $1\times 10^6$ documents.
On the other hand, the memory usage of \textsf{Bunker-B} keeps increasing when we delete more documents as it should maintain the deleted keyword-document pairs on the server.
Within the enclave, \textsf{Bunker-B} does not maintain any persistent data structure while \textsf{SGX-SE1} and \textsf{SGX-SE2} need to store the necessary information for deletion.
For \textsf{SGX-SE1}, it caches all the document \textit{id} in the enclave, which leads to notably high memory usage (e.g., $304$ MB when deleting 25\% documents, and $355$ MB when deleting 75\%).
The memory resource requests in \textsf{SGX-SE1} triggers the paging mechanism of the SGX, resulting in a larger query delay as presented above.
\textsf{SGX-SE2} successfully prevents the paging by using the Bloom filter.
After applying a Bloom filter with the false positive rate $10^{-4}$, \textsf{SGX-SE2} only needs $34$ MB to store all keyword-document pairs ($1.18\times 10^7$ pairs) and maintains a low query delay over the dataset.

\vspace{-5pt}
\subsection{Performance evaluation on the Enron dataset}\label{subsec:enron}
We use a real world dataset to illustrate the practicality of the proposed scheme.
Since the bulk deletion (e.g. delete 50\%) is rare in real world, we only focus on the setting with a small deletion portion.
Therefore, in the following experiments, we insert the whole Enron dataset and test the average runtime for insertion/deletion as well as the query delay with a small deletion portion (25\%).

\begin{table}[!t]
	\caption{Average time ($\mu$s) for adding a keyword-doc pair from Enron dataset and removing 25\% documents under different schemes.}
	\label{tlb:enron_setup}
	\centering
	\begin{tabular}{|c|c|c|c|}
		\hline        		
		Operation & BunkerB & SGX-SE1 & SGX-SE2 \\
		\hline
		Insertion & $12$ & $7$ & $8$ \\
		\hline
		Deletion (25\%, 129,305 documents) & $12$ & $4$ & $4$ \\
		\hline
	 	\end{tabular}
	 	
\end{table}

\noindent \textbf{Insertion and deletion}: As described in Section~\ref{subsec:synthesis}, our schemes are more efficient for the insertion and deletion if the number of keyword-document pairs is larger than the number of documents.
The evaluation result on the Enron dataset further verifies our observation: as shown in Table~\ref{tlb:enron_setup}, our schemes only need $7\ \mu$s and $8\ \mu$s respectively to insert one keyword-document pair while \textsf{Bunker-B} needs $12\ \mu$s to do that.
Besides, both of \textsf{SGX-SE1} and \textsf{SGX-SE2} only takes $4\ \mu$s to delete one document, but \textsf{Bunker-B} still requires $12\ \mu$s to execute the same algorithm as the insertion.

\begin{figure}
	\centering
	\includegraphics[width=0.35\textwidth,height=2.8cm]{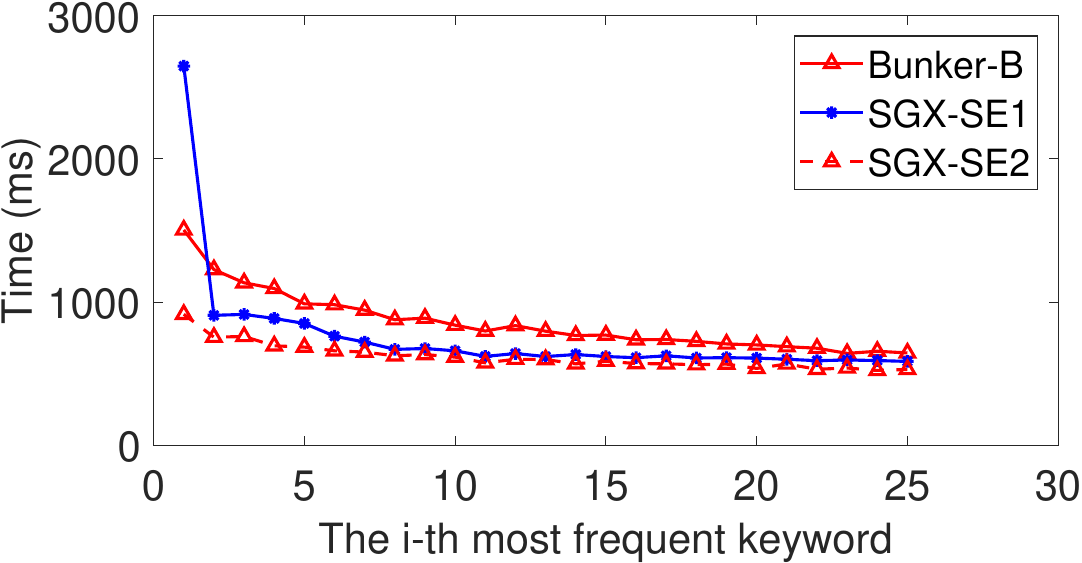}
	\caption{The query delay of querying the i-th most frequent keyword in the Enron dataset under different schemes (insert all documents and delete 25\% of them).}\label{fig:enron_25}
\end{figure}

\noindent \textbf{Query delay}: Finally, we present the query delay when using the Enron dataset.
As the Enron dataset has more keyword-document pairs than our synthesis dataset, deleting 25\% documents still triggers paging, as it includes more keyword-document pairs than the whole synthesis dataset.
In Fig.~\ref{fig:enron_25}, we present the query delay when querying the top-25 frequent keywords in the Enron dataset.
The result shows that \textsf{SGX-SE2} maintains a relative low query delay ($530$ ms to $900$ ms) while \textsf{SGX-SE1} needs $580$ ms to $2.6$ s and \textsf{Bunker-B} requires $645$ ms to $1.5$ s.
This above result further illustrates that \textsf{SGX-SE2} can both prevent the paging within the SGX enclave and eliminate the cost of re-encryption.
Note that using \textit{exit-less} system calls can further improve the performance in the enclave by eliminating the overheads of enclave exits and optimising \textit{enclave page cache} accesses~\cite{Orenbach17}. Hence, we will evaluate this performance in the future work.

\vspace{-10pt}
\section{Conclusion}
\label{sec:con}
In this paper, we leverage the advance of Intel SGX to design and implement forward and backward private dynamic searchable encryption schemes. We carefully analyse the limitations of the recent theoretical constructions and propose new designs to avoid the bottleneck of the SGX enclave. We present a basic scheme and then further optimise it for better performance. We implement prior work and our schemes, and conduct a detailed performance comparison. The results show that our designs are more efficient in query latency and data deletion.

\section*{Acknowledgement}
This work was supported by the Australian Research Council (ARC) Discovery Project grant DP200103308. 
\vspace{-10pt}
%Our two other schemes improve from the basic.

%The 1st optimised eliminate the bottleneck in loading docs into enclave by using bloom filter. 

%But bloomfilter is still globally user for all keywords, depending vectorsize of bloom.

%Therefore, third scheme uses state management to swap memory between enclave and outside in an encryption data structure manner

%BunkerB keep evrything outside, but it is not practice in streaming IoT environment.
%BunkerB has linear time consumpyion with keyword frequency for sesrch/update latency.
%It does not decay time search like our protocol 
%if we search for multiple keywords.
%Our schemes improve searching if we search same keyword again.

\bibliographystyle{IEEEtranS}
\bibliography{sgxse}

% Generated by IEEEtranS.bst, version: 1.12 (2007/01/11)
\begin{thebibliography}{10}
\providecommand{\url}[1]{#1}
\csname url@samestyle\endcsname
\providecommand{\newblock}{\relax}
\providecommand{\bibinfo}[2]{#2}
\providecommand{\BIBentrySTDinterwordspacing}{\spaceskip=0pt\relax}
\providecommand{\BIBentryALTinterwordstretchfactor}{4}
\providecommand{\BIBentryALTinterwordspacing}{\spaceskip=\fontdimen2\font plus
\BIBentryALTinterwordstretchfactor\fontdimen3\font minus
  \fontdimen4\font\relax}
\providecommand{\BIBforeignlanguage}[2]{{%
\expandafter\ifx\csname l@#1\endcsname\relax
\typeout{** WARNING: IEEEtranS.bst: No hyphenation pattern has been}%
\typeout{** loaded for the language `#1'. Using the pattern for}%
\typeout{** the default language instead.}%
\else
\language=\csname l@#1\endcsname
\fi
#2}}
\providecommand{\BIBdecl}{\relax}
\BIBdecl

\bibitem{Amjad19}
G.~Amjad, S.~Kamara, and T.~Moataz, ``{Forward and Backward Private Searchable
  Encryption with SGX},'' in \emph{{EuroSec}'19}.

\bibitem{Bindschaedler18}
V.~Bindschaedler, P.~Grubbs, D.~Cash, T.~Ristenpart, and V.~Shmatikov, ``{The
  Tao of Inference in Privacy-protected Databases},'' \emph{Proc. VLDB Endow.},
  2018.

\bibitem{Borges18}
G.~Borges, H.~Domingos, B.~Ferreira, J.~Leitão, T.~Oliveira, and B.~Portela,
  ``{BISEN: Efficient Boolean Searchable Symmetric Encryption with
  Verifiability and Minimal Leakage},'' in \emph{{SRDS}'19}.

\bibitem{Bost16}
R.~Bost, ``{Sophos - Forward Secure Searchable Encryption},'' in \emph{{ACM
  CCS}'16}.

\bibitem{RaphaelBO17}
R.~Bost, B.~Minaud, and O.~Ohrimenko, ``{Forward and Backward Private
  Searchable Encryption from Constrained Cryptographic Primitives},'' in
  \emph{{ACM CCS}'17}, 2017.

\bibitem{Brasser19}
F.~Brasser, S.~Capkun, A.~Dmitrienko, T.~Frassetto, K.~Kostiainen, and A.-R.
  Sadeghi, ``{DR.SGX: Automated and Adjustable Side-Channel Protection for SGX
  using Data Location Randomization},'' in \emph{{ACSAC}'19}.

\bibitem{Brasser17}
F.~Brasser, U.~M\"{u}ller, A.~Dmitrienko, K.~Kostiainen, S.~Capkun, and A.-R.
  Sadeghi, ``{Software Grand Exposure: SGX Cache Attacks Are Practical},'' in
  \emph{{WOOT}'17}.

\bibitem{CashGP15}
D.~Cash, P.~Grubbs, J.~Perry, and T.~Ristenpart, ``{Leakage-Abuse Attacks
  against Searchable Encryption},'' in \emph{{ACM CCS}'15}, 2015.

\bibitem{Cash14}
D.~Cash, J.~Jaeger, S.~Jarecki, and C.~Jutla, ``{Dynamic Searchable Encryption
  in Very Large Databases: Data Structures and Implementation},'' in
  \emph{{NDSS}'14}.

\bibitem{Priebe18}
P.~Christian, V.~Kapil, and C.~Manuel, ``{EnclaveDB: A Secure Database using
  SGX},'' in \emph{{IEEE S\&P}'18}.

\bibitem{CurtmolaGKO06}
R.~Curtmola, J.~Garay, S.~Kamara, and R.~Ostrovsky, ``{Searchable Symmetric
  Eencryption: Improved Definitions and Efficient Constructions},'' in
  \emph{{ACM CCS}'06}.

\bibitem{Huayi19}
H.~Duan, C.~Wang, X.~Yuan, Y.~Zhou, Q.~Wang, and K.~Ren, ``{LightBox:
  Full-stack Protected Stateful Middlebox at Lightning Speed},'' in \emph{{ACM
  CCS}'19}.

\bibitem{Fuhry17}
B.~Fuhry, R.~Bahmani, F.~Brasser, F.~Hahn, F.~Kerschbaum, and A.~Sadeghi,
  ``{HardIDX: Practical and Secure Index with {SGX}},'' in \emph{{DBSec}'17}.

\bibitem{GharehChamani18}
J.~Ghareh~Chamani, D.~Papadopoulos, C.~Papamanthou, and R.~Jalili, ``{New
  Constructions for Forward and Backward Private Symmetric Searchable
  Encryption},'' in \emph{{ACM CCS}'18}.

\bibitem{Kamara12}
S.~Kamara, C.~Papamanthou, and T.~Roeder, ``{Dynamic Searchable Symmetric
  Encryption},'' in \emph{{ACM CCS}'12}, 2012.

\bibitem{LaiPSL18}
S.~Lai, S.~Patranabis, A.~Sakzad, J.~K. Liu, D.~Mukhopadhyay, R.~Steinfeld
  \emph{et~al.}, ``Result pattern hiding searchable encryption for conjunctive
  queries,'' in \emph{{ACM CCS}'18.}, 2018.

\bibitem{Mishra18}
P.~{Mishra}, R.~{Poddar}, J.~{Chen}, A.~{Chiesa}, and R.~A. {Popa}, ``{Oblix:
  An Efficient Oblivious Search Index},'' in \emph{{IEEE S\&P}'18}.

\bibitem{Orenbach17}
M.~Orenbach, P.~Lifshits, M.~Minkin, and M.~Silberstein, ``Eleos: Exitless os
  services for sgx enclaves,'' ser. EuroSys ’17.

\bibitem{Ren20}
K.~Ren, Y.~Guo, L.~Jiaqi, X.~Jia, C.~Wang, Y.~Zhou, S.~Wang, N.~Cao, and F.~Li,
  ``Hybridx: New hybrid index for volume-hiding range queries in data
  outsourcing services,'' in \emph{{ICDCS}'20}.

\bibitem{Shinde16}
S.~Shinde, Z.~L. Chua, V.~Narayanan, and P.~Saxena, ``{Preventing Page Faults
  from Telling Your Secrets},'' in \emph{{ACM AsiaCCS}'16}.

\bibitem{SoWa00}
D.~Song, D.~Wagner, and A.~Perrig, ``{Practical Techniques for Searches on
  Encrypted Data},'' in \emph{{IEEE S\&P}'00}.

\bibitem{StefanovPS14}
E.~Stefanov, C.~Papamanthou, and E.~Shi, ``{Practical Dynamic Searchable
  Symmetric Encryption with Small Leakage},'' in \emph{{NDSS}'14}, 2014.

\bibitem{SunYLS18}
S.-F. Sun, X.~Yuan, J.~Liu, R.~Steinfeld, A.~Sakzad, V.~Vo \emph{et~al.},
  ``{Practical Backward-Secure Searchable Encryption from Symmetric Puncturable
  Encryption},'' in \emph{{ACM CCS}'18}, 2018.

\bibitem{Yarom14}
Y.~Yarom and K.~Falkner, ``{FLUSH+RELOAD: A High Resolution, Low Noise, L3
  Cache Side-Channel Attack},'' in \emph{{USENIX Security}'14}.

\bibitem{ZhangKP16}
Y.~Zhang, J.~Katz, and C.~Papamanthou, ``{All Your Queries Are Belong to Us:
  The Power of File-Injection Attacks on Searchable Encryption},'' in
  \emph{{USENIX Security}'16}.

\bibitem{Zuo19}
C.~Zuo, S.-F. Sun, J.~Liu, J.~Shao, and J.~Pieprzyk, ``Dynamic searchable
  symmetric encryption with forward and stronger backward privacy,'' in
  \emph{{ESORICS}'19}.

\bibitem{Zuo18}
C.~Zuo, S.-F. Sun, J.~K. Liu, J.~Shao, and J.~Pieprzyk, ``Dynamic searchable
  symmetric encryption schemes supporting range queries with forward (and
  backward) security,'' in \emph{{ESORICS}'18}.

\end{thebibliography}
\end{document}